\documentclass{IEEEoj}
\usepackage[T1]{fontenc}
\usepackage[utf8]{inputenc}
\usepackage{cite}
\usepackage{amsmath,amssymb,amsfonts}
\usepackage{algorithm}
\usepackage{algpseudocodex}
\usepackage{graphicx,color}
\usepackage{textcomp}
\usepackage{booktabs}
\usepackage{multirow}
\usepackage{array}
\usepackage{tabularx}
\usepackage{url}
\usepackage{float}
\usepackage{needspace}
\usepackage{comment}
\usepackage{acro}
\usepackage{longtable}
\DeclareAcronym{qpp}{
    short = QPP,
    long = {Quality, Priority, and Pre-emption}
}

\DeclareAcronym{lmr}{
    short = LMR,
    long = Land Mobile Radio,
}

\DeclareAcronym{leo}{
    short = LEO,
    long = Low Earth Orbit,
}

\DeclareAcronym{tetra}{
    short = TETRA,
    long = Terrestrial Trunked Radio,
}

\DeclareAcronym{dmr}{
    short = DMR,
    long = Digital Mobile Radio,
}

\DeclareAcronym{p25}{
    short = P25,
    long = Project 25,
}

\DeclareAcronym{mcx}{
    short = MCX,
    long =  Mission Critical Services,
}

\DeclareAcronym{mcptt}{
    short = MCPTT,
    long =  Mission Critical Push-to-Talk,
}

\DeclareAcronym{ptt}{
    short = PTT,
    long =  Push-to-Talk,
}

\DeclareAcronym{mcvideo}{
    short = MCVideo,
    long =  Mission Critical Video,
}

\DeclareAcronym{mcdata}{
    short = MCData,
    long =  Mission Critical Data,
}

\DeclareAcronym{uav}{
    short = UAV,
    long =  Unmanned Aerial Vehicle,
}

\DeclareAcronym{ugv}{
    short = UGV,
    long =  Unmanned Ground Vehicle,
}

\DeclareAcronym{mec}{
    short = MEC,
    long =  Multi-access Edge Computing,
}

\DeclareAcronym{ntn}{
    short = NTN,
    long =  Non-Terrestrial Networks,
}

\DeclareAcronym{5gc}{
    short = 5GC,
    long =  5G Core,
}

\DeclareAcronym{nr}{
    short = NR,
    long =  New Radio,
}

\DeclareAcronym{qos}{
    short = QoS,
    long =  Quality of Service,
}

\DeclareAcronym{qoe}{
    short = QoE,
    long =  Quality of Experience,
}

\DeclareAcronym{urllc}{
    short = URLLC,
    long =  Ultra-Reliable Low-Latency Communication ,
}

\DeclareAcronym{sip}{
    short = SIP,
    long =  Session Initiation Protocol ,
}

\DeclareAcronym{3gpp}{
    short = 3GPP,
    long =  3rd Generation Partnership Project,
}

\DeclareAcronym{as}{
    short = AS,
    long =  Application Server,
}

\DeclareAcronym{lte}{
    short = LTE,
    long =  Long-Term Evolution,
}

\DeclareAcronym{kpi}{
    short = KPI,
    long =  Key Performance Indicator,
}

\DeclareAcronym{nsa}{
    short = NSA,
    long =  Non-Standalone,
}

\DeclareAcronym{sa}{
    short = SA,
    long =  Standalone,
}

\DeclareAcronym{vonr}{
    short = VoNR,
    long =  Voice Over New Radio,
}

\DeclareAcronym{ims}{
    short = IMS,
    long =  IP Multimedia Subsystem,
}

\DeclareAcronym{frmcs}{
    short = FRMCS,
    long =  Future Railway Mobile Communication System,
}

\DeclareAcronym{ran}{
    short = RAN,
    long =  Radio Access Network,
}

\DeclareAcronym{ppdr}{
    short = PPDR,
    long =  Public Protection and Disaster Relief,
}

\DeclareAcronym{b5g}{
    short = B5G,
    long =  Beyond 5G,
}

\DeclareAcronym{b2bua}{
    short = BTBUA,
    long =  Back-to-Back User Agent,
}

\DeclareAcronym{5qi}{
    short = 5QI,
    long =  5G QoS Identifier,
}

\DeclareAcronym{qfi}{
    short = QFI,
    long =  QoS Flow Identifier,
}

\DeclareAcronym{pdu}{
    short = PDU,
    long =  Protocol Data Unit,
}

\DeclareAcronym{ue}{
    short = UE,
    long =  User Equipment,
}

\DeclareAcronym{upf}{
    short = UPF,
    long =  User Plane Function,
}

\DeclareAcronym{pdb}{
    short = PDB,
    long =  Packet Delay Budget,
}

\DeclareAcronym{per}{
    short = PER,
    long =  Packet Error Rate,
}

\DeclareAcronym{mdbv}{
    short = MDBV,
    long =  Maximum Data Burst Volume,
}

\DeclareAcronym{gbr}{
    short = GBR,
    long =  Guaranteed Bit Rate,
}

\DeclareAcronym{mbr}{
    short = MBR,
    long =  Maximum Bit Rate,
}

\DeclareAcronym{mfbr}{
    short = MFBR,
    long =  Maximum Flow Bit Rate,
}

\DeclareAcronym{arp}{
    short = ARP,
    long =  Allocation and Retention Priority,
}

\DeclareAcronym{gnb}{
    short = gNB,
    long =  Next-Generation NodeB,
}

\DeclareAcronym{mc}{
    short = MC,
    long =  Mission-Critical,
}

\DeclareAcronym{csc}{
    short = CSC,
    long =  Common Services Core,
}

\DeclareAcronym{amf}{
    short = AMF,
    long =  Access and Mobility Management Function,
}

\DeclareAcronym{smf}{
    short = SMF,
    long =  Session Management Function,
}

\DeclareAcronym{udm}{
    short = UDM,
    long =  Unified Data Management,
}

\DeclareAcronym{pcf}{
    short = PCF,
    long =  Policy Control Function,
}

\DeclareAcronym{ausf}{
    short = AUSF,
    long =  Authentication Server Function,
}

\DeclareAcronym{rtp}{
    short = RTP,
    long =  Real-time Transport Protocol,
}

\DeclareAcronym{sdp}{
    short = SDP,
    long =  Session Description Protocol,
}

\DeclareAcronym{pcscf}{
    short = P-CSCF,
    long =  Proxy Call Session Control Function,
}

\DeclareAcronym{icscf}{
    short = I-CSCF,
    long =  Interrogating Call Session Control Function,
}

\DeclareAcronym{scscf}{
    short = S-CSCF,
    long =  Serving Call Session Control Function,
}

\DeclareAcronym{gms}{
    short = GMS,
    long =  Group Management Service,
}

\DeclareAcronym{nwdaf}{
  short = NWDAF,
  long = Network Data Analytics Function
}

\DeclareAcronym{af}{
  short = AF,
  long = Application Function
}

\DeclareAcronym{lbo}{
  short = LBO,
  long = Local Breakout
}

\DeclareAcronym{hr}{
  short = HR,
  long = Home Routed
}

\DeclareAcronym{plmn}{
  short = PLMN,
  long = Public Land Mobile Network
}

\DeclareAcronym{mbs}{
  short = MBS,
  long = Multicast-Broadcast Services
}

\DeclareAcronym{idmf}{
  short = IDMF,
  long = Identity Management Function
}

\DeclareAcronym{cms}{
  short = CMS,
  long = Configuration Management Service
}

\DeclareAcronym{lms}{
  short = LMS,
  long = Location Management Server 
}

\DeclareAcronym{mos}{
  short = MOS,
  long = Mean Opinion Score
}

\DeclareAcronym{ngap}{
  short = NGAP,
  long = NG Application Protocol,
}

\DeclareAcronym{dcs}{
  short = DCS,
  long = Data Content Server,
}

\DeclareAcronym{pc}{
    short = PC,
    long =  Pre-emption Capability,
}

\DeclareAcronym{pv}{
    short = PV,
    long =  Pre-emption Vulnerability,
}

\DeclareAcronym{d2d}{
    short = D2D,
    long =  Device-to-device,
}

\DeclareAcronym{atsss}{
    short = ATSSS,
    long =  {Access Traffic Steering, Switching, and Splitting},
}

\DeclareAcronym{mptcp}{
    short = MPTCP,
    long =  Multipath TCP,
}
\usepackage[hidelinks]{hyperref}
\usepackage{tikz}
\usetikzlibrary{positioning,arrows.meta}
\usepackage[normalem]{ulem}
\usepackage{xcolor}

\def\BibTeX{{\rm B\kern-.05em{\sc i\kern-.025em b}\kern-.08em
    T\kern-.1667em\lower.7ex\hbox{E}\kern-.125emX}}
\AtBeginDocument{\definecolor{ojcolor}{cmyk}{0.93,0.59,0.15,0.02}}
\def\OJlogo{\vspace{-4pt}\hskip-4pt\includegraphics[height=18pt]{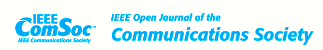}}

\begin{document}
\receiveddate{XX Month, XXXX}
\reviseddate{XX Month, XXXX}
\accepteddate{XX Month, XXXX}
\publisheddate{XX Month, XXXX}
\currentdate{18 June, 2026}
\doiinfo{OJCOMS.XXXX.XXXXXX}


\title{QPP-Aware MCX over Tactical Edge 3D Networks in Public-Safety Communications}

\author{José Oliveira\IEEEauthorrefmark{1}\IEEEauthorrefmark{2}, Pedro Valente\IEEEauthorrefmark{1}\IEEEauthorrefmark{2},
Tiago Mendes\IEEEauthorrefmark{1}\IEEEauthorrefmark{2}, André Vasconcelos\IEEEauthorrefmark{1}\IEEEauthorrefmark{2}, \\ Tiago Fernandes\IEEEauthorrefmark{1}\IEEEauthorrefmark{2}, Rhyan Matos\IEEEauthorrefmark{1}\IEEEauthorrefmark{2}, Duarte Raposo\IEEEauthorrefmark{1}, \\ Pedro Rito\IEEEauthorrefmark{1}\IEEEauthorrefmark{2},
Marco Araújo\IEEEauthorrefmark{1}\IEEEauthorrefmark{3}, Susana Sargento\IEEEauthorrefmark{1}\IEEEauthorrefmark{2}}
\affil{Instituto de Telecomunicacoes, Aveiro, 3810-193, Portugal}
\affil{Universidade de Aveiro, Aveiro, 3810-193, Portugal}
\affil{Faculdade de Engenharia da Universidade do Porto, Porto, 4200-465, Portugal}
\corresp{CORRESPONDING AUTHOR: oliveira.jose@ua.pt.}
\markboth{}{Oliveira \textit{et al.}}

\begin{abstract}

During disasters, public mobile networks may become unavailable, congested, or degraded, requiring deployable communication systems to support emergency operations. This article proposes a mission-critical communications system supporting \ac{mcptt}, \ac{mcvideo}, and \ac{mcdata} over a private 5G standalone network with \ac{ntn} connectivity. A context-aware \ac{qpp} mechanism is proposed to dynamically adapt network resource allocation according to the operational state of \ac{mcx}. Extensions to the Open5GS \ac{pcf} and \ac{smf} enable priority and pre-emption updates for existing flows without session re-establishment. The architecture supports tactical edge deployment of service and media functions while retaining centralized service control. Experimental evaluation considers three configurations: terrestrial backhaul with centralized media processing, satellite backhaul with centralized media processing, and satellite backhaul with edge media processing. Results show that tactical edge processing reduces communication delay and packet loss. Under satellite backhaul, edge processing reduced the 95th-percentile \ac{mcptt} audio transport delay by 60.03 ms and packet loss by 9.42 percentage points, while achieving a 95th-percentile \ac{mcptt} access time of 184.52 ms. These findings demonstrate the benefits of tactical edge media processing and provide an experimental foundation for context-aware priority mechanisms in public-safety networks integrating terrestrial and non-terrestrial components.

\end{abstract}


\maketitle

\section{Introduction}

The increasing frequency and intensity of weather and climate-related disasters are placing growing pressure on public safety and emergency response systems worldwide~\cite{disasters}. Large-scale events, such as floods, wildfires, and severe storms, can simultaneously increase demand for emergency services, damage terrestrial communication infrastructure, and disrupt coordination between response agencies \cite{hazra2020postdisaster, al_turjman2022postdisaster}. These conditions motivate resilient communication systems that can be rapidly deployed, provide controlled service quality for mission-critical applications, and remain operational when conventional infrastructure is degraded or unavailable.


Historically, public safety communications have relied on \acp{lmr}, including technologies such as \ac{tetra}, \ac{p25}, and \ac{dmr}. Designed primarily for mission-critical voice, these systems provide essential capabilities such as group communication, direct-mode operation, and push-to-talk, but offer limited support for broadband data and multimedia services \cite{michalopoulou2021_5g}. Consequently, they are poorly suited to emerging robotic operations involving \acp{uav} and ground rovers, which require reliable exchange of high-rate sensor data, real-time video, telemetry, and control information. These requirements motivate the transition toward broadband public safety networks based on 5G and future 6G technologies.

Within this transition, the \ac{3gpp} \ac{mcx} framework standardizes mission-critical voice, video, and data services over cellular networks. \ac{mcptt} provides individual and group voice communication with priority and floor control, while \ac{mcvideo} and \ac{mcdata} support real-time video, situational awareness, messaging, positioning, and telemetry~\cite{3gpp_mc_common}. Supporting these services concurrently requires differentiated network treatment, as latency-sensitive voice and signaling must coexist with high-rate video and background traffic. In 5G, \ac{qos} flows and policy control mechanisms across the \ac{5gc} and \ac{nr} enable service differentiation according to communication requirements and operational priority, here referred to as \acf{qpp} \cite{tcca_qpp}. Together with capabilities for group communication, extended coverage, and positioning, these mechanisms make 5G \ac{nr} a key enabler for broadband mission-critical communications~\cite{mcx5g}.

During large-scale disasters, terrestrial communication infrastructure may become damaged, congested, or unavailable, limiting network coverage and capacity when communication demand is high. Deployable 5G networks can restore local connectivity through portable infrastructure and aerial radio nodes, while \ac{ntn} links can provide backhaul when terrestrial connectivity is disrupted~\cite{deployable_5g,3d}. In this work, the integration of terrestrial, aerial, and \ac{ntn} components is referred to as a 3D public-safety network.
The heterogeneous latency and capacity of these links make function placement an important architectural consideration. Deploying latency-sensitive user plane and \ac{mcx} media functions at the edge can keep local traffic within the deployable network, reducing dependence on remote backhaul while maintaining connectivity to external services when backhaul resources permit~\cite{11008665}.

Despite these advances, end-to-end integration of mission-critical services with deployable 5G architectures remains insufficiently explored. In particular, a gap persists between \ac{mcx} service priorities and their dynamic enforcement through \acf{qpp} mechanisms across the \ac{5gc} and \ac{ran}~\cite{tcca_qpp}. Furthermore, the impact of edge function placement on \ac{mcx} signaling and media traffic over constrained \ac{ntn} backhaul remains insufficiently characterized in experimental deployments.

This work investigates the dynamic translation of \ac{mcx} service priorities into \ac{qpp} enforcement across the \ac{5gc} and \ac{ran}, and evaluates how \ac{mcx} function placement affects signaling and media traffic over constrained \ac{ntn} backhaul. The main contributions of this work are:

\begin{itemize}
    
\item \textbf{A deployable multi-service \ac{mcx} architecture} integrating \ac{mcptt}, \ac{mcvideo}, and \ac{mcdata} over a private 5G \ac{sa} and \ac{ims} infrastructure. The architecture combines common \ac{mcx} support functions, service specific \ac{sip} \ac{as} modules, edge capable media processing, tactical \ac{uav} endpoints and \ac{ntn} backhaul within a common experimental platform.

\item \textbf{A cross-layer \ac{qpp} model for \ac{mcx}} that maps service and operational priorities to 5G network treatment. \ac{5qi} defines service level forwarding requirements, while dynamic operational priority determines \ac{arp}-based admission and preemption, complementing application-level \ac{mcptt} floor control.

\item \textbf{Dynamic \ac{qpp} enforcement for active \ac{mcx} communications.} Extensions upstream to Open5GS map the N5 \texttt{resPrio} attribute to configurable \ac{arp} parameters, and dynamically update the priority and preemption treatment of established \ac{qos} flows. This enables operational transitions, such as emergency \ac{mcptt}, without session re-establishment or modification of the baseline \ac{5qi}.

\item \textbf{A persistent unicast media architecture for \ac{mcptt} and \ac{mcvideo}.} The design maintains stable per-participant signaling and media state, and allows \ac{mcptt} floor transitions to modify media forwarding without reconstructing the underlying session or network treatment.

\item \textbf{An end-to-end experimental evaluation across service, network, and deployment conditions.} Measurements across the \ac{mcx}, \ac{sip}/\ac{ims}, media, 5G Core, and \ac{ran} layers are correlated under concurrent service load, resource scarcity and dynamic priority transitions. The evaluation further investigates an aerial edge deployment in which the \ac{gnb}, \ac{upf} and RTP engine are placed at the \ac{uav} edge during \ac{ntn} backhaul, allowing local mission-critical media to remain at the edge while signaling goes through the non-terrestrial backhaul.
\end{itemize}

The remainder of this article is organized as follows. Section~\ref{sec:related_work} reviews related work on mission-critical communications and enabling 5G, edge, aerial, and non-terrestrial technologies. Section~\ref{sec:use_cases} presents representative \ac{ppdr} scenarios, and derives the corresponding system requirements. Section~\ref{sec:cross_layer_priority} introduces the proposed cross-layer \ac{mcx} service and \ac{qos} model for dynamic network priority control. Section~\ref{sec:architecture} describes the proposed architecture and its implementation across the \ac{mcx}, \ac{ims}, 5G, edge, and non-terrestrial domains. Section~\ref{sec:experimental_validation} presents the experimental methodology and evaluates service performance, dynamic priority, edge placement, and \ac{ntn} backhaul. Section~\ref{sec:conclusions} concludes the article and outlines future work.

\section{Background and Related Work}
\label{sec:related_work}

This section reviews the foundations and evolution of \ac{3gpp} \ac{mcx} toward 5G-Advanced and 6G enabled public-safety communication systems. It first summarizes the main architectural principles of standardized mission-critical services, and reviews experimental \ac{mcx} implementations addressing service performance, scalability, and network \ac{qos}. It then discusses enabling technologies for future mission-critical systems, including edge computing, network slicing, intelligent orchestration, \acp{uav}, and terrestrial and non-terrestrial networks.

\subsection{3GPP Mission-Critical Services and Implementations}

A broader perspective on the role of 5G \ac{nr} in public safety communications is provided in \cite{mcx5g}. The authors identify requirements including ubiquitous connectivity, efficient group communication, extended coverage, accurate positioning, sidelink operation, support for isolated or deployable networks and prioritization of mission-critical traffic. Their work positions 5G \ac{nr} as an important technological basis for complementing traditional \acp{lmr} and extending public-safety communication toward broadband multimedia services.


Research on mission-critical communications has progressively moved from standardized architectures and analytical studies toward experimental platforms capable of evaluating service behavior over broadband cellular networks. The authors in \cite{pttapp} extend the MCOP framework with a Next-Generation \ac{mcptt} architecture that exposes packetized voice streams directly to the application layer. This enables mechanisms such as store-and-replay, asynchronous \ac{ptt} concurrent conversations, application-level message prioritization, speech-to-text and text-to-speech processing, and device-to-device store and forward communication. Rather than focusing exclusively on conventional \ac{mcptt} access or mouth-to-ear latency, the work investigates how additional application-level control over voice streams can improve responder interaction. However, the approach remains primarily centered on \ac{mcptt} and application layer voice manipulation, without integrating the broader \ac{mcx} service set or coordinating application-level operational state with 5G Core and \ac{ran} \ac{qos} mechanisms.

The work in \cite{lteptt} investigates the use of \ac{lte} and \ac{mcptt} as a converged communication infrastructure for airport operations, and proposes an interworking function between \ac{mcptt} and ED-137-based air-traffic-control radio systems. The proposed approach exploits \ac{mcptt} floor priority and Floor Revoke procedures to preserve the precedence of air-traffic-control communications, while \ac{lte} dedicated bearers and policy control mechanisms provide differentiated network treatment. The work illustrates the importance of coordinating service level communication priority with broadband network \ac{qos}. However, its scope is restricted to a specific aviation interworking scenario based on \ac{lte}, and it does not investigate the dynamic modification of 5G \ac{qos}-Flow priority in response to operational state transitions within an already established mission-critical session.

\ac{d2d} communication has also been investigated as a mechanism for maintaining mission-critical connectivity when conventional cellular infrastructure is unavailable. The work in \cite{d2d} analyzes several network-controlled and device-controlled \ac{d2d} configurations for \ac{mcptt}, and compares direct and cellular transmission under fading, path loss, and co-channel interference. The study is primarily focused on radio-link behavior rather than on the integration of off-network operation into a complete \ac{mcx} architecture.

The impact of the evolution from \ac{lte} toward 5G architectures on \ac{mcptt} performance is analyzed in \cite{mcptt_5gnsa}. The authors examine \ac{mcptt} signaling and media procedures, and decompose the contribution of different network segments to \ac{3gpp}-defined \acp{kpi}. \ac{lte}, \ac{mec}-assisted \ac{lte}, and 5G \ac{nsa} architectures are compared, showing that edge placement can improve latency-sensitive performance, particularly mouth-to-ear delay. The results also indicate that satisfying strict \ac{mcptt} requirements remains challenging when the standardized service chain is considered end to end, reinforcing the importance of service placement and transport-network design.

Capacity and scalability aspects are further investigated in \cite{ran_limitations}, where the authors evaluate radio-access limitations for mission-critical group voice over \ac{lte} and 5G. Their analysis demonstrates that large group communication can remain constrained by \ac{ran} capacity and by protocol mechanisms involved in scheduling and identifying downlink traffic. This is particularly relevant in public-safety incidents where many responders may simultaneously participate in the same operational group. The study provides a more detailed capacity characterization than estimations based only on peak cell throughput, but does not investigate how group-control logic, media distribution, and network-resource management interact in an end-to-end \ac{mcx} implementation.

Several works move beyond analytical evaluation toward practical \ac{mcx} experimentation. The proof-of-concept implementation in \cite{open_mcptt} integrates open-source components and an \ac{mcptt} Application Server to realize \ac{mcptt} services over a private 5G network. The platform is evaluated using \ac{3gpp} oriented metrics including \ac{mcptt} access time, end-to-end access time, and mouth-to-ear latency, demonstrating both the feasibility and current performance limitations of open \ac{mcptt} implementations. However, the implementation primarily targets \ac{mcptt}, and its evaluation focuses on service functionality and end-to-end performance rather than on dynamically modifying the network treatment of an active communication in response to changes in \ac{mcx} operational state. 

Open-source \ac{vonr} prototypes have additionally demonstrated interoperable \ac{ims}-based voice operation over open 5G \ac{sa} networks using architectural components relevant to \ac{mcx} deployments~\cite{open_vonr}. However, conventional conversational voice does not provide mission-critical group communication, floor control, service level priority, or emergency preemption procedures.

Field experimentation with \ac{mcx} is addressed in \cite{frmcs}, where mission-critical services are evaluated in the context of the \ac{frmcs}. The authors report early field trials over 5G including \ac{mcptt} point-to-point and group calls, emergency calls, \ac{mcdata}/IPConn services, and \ac{qos} measurements under stationary and mobility conditions. The contribution demonstrates the importance of evaluating mission-critical communication under realistic operational conditions, including mobility, handover, and end-to-end service establishment. Importantly, the authors identify traffic steering, decentralized media-plane operation, and the analysis of \ac{qos} parameters and their dependencies across protocol layers as relevant directions for future investigation. These directions are closely related to the questions addressed in this work, particularly the dynamic association between mission-critical operational state, network treatment and the placement of media functions closer to the access network.


Collectively, these works demonstrate substantial progress in \ac{mcx} implementation, performance characterization, and the use of differentiated mobile-network \ac{qos}. However, experimental platforms remain predominantly focused on individual services, particularly \ac{mcptt}, and different levels of application–network integration should be distinguished. Configuring dedicated bearers or \ac{qos} profiles for mission-critical traffic establishes service-aware treatment, while policy-authorization mechanisms can additionally allow an application function to request network resources. Comparatively less attention has been given to dynamic transitions occurring within an already active \ac{mcx} communication, where changes in application state must modify network priority without reconstructing the service session or its \ac{qos} Flow.

\subsection{5G-Advanced and 6G Enablers for Mission-Critical Communications}

The evolution toward 5G-Advanced and 6G introduces additional mechanisms that may improve the flexibility, resilience, and autonomy of mission-critical communication systems. Relevant directions include advanced network slicing, edge and cloud-native service deployment, AI-assisted network management, integrated terrestrial and non-terrestrial connectivity, and increasingly dynamic placement of communication and computing resources \cite{habibi_6g,nguyen_6g_ntn}. For deployable public-safety networks, edge computing and \ac{ntn} integration are particularly relevant because they can reduce dependence on remote infrastructure while extending connectivity beyond the coverage of surviving terrestrial networks.

A network-slicing architecture for mission-critical communications is presented in \cite{slicing}. The authors deploy a commercial mission-critical services product in a fully virtualized, multi-domain network-slicing environment compliant with ETSI NFV-MANO. The platform demonstrates automated deployment, slice isolation, differentiated performance, and management flexibility, illustrating how virtualization and network slicing can provide dedicated network resources for mission-critical services. The resulting differentiation is primarily established through slice-level resource allocation and isolation. In contrast, the problem addressed in this work concerns differentiation among individual mission-critical communications within the available network resources. Service-specific \ac{5qi} profiles define their baseline forwarding characteristics, while \ac{arp} and its preemption attributes are dynamically modified according to the operational state of an active \ac{mcx} communication. The two approaches therefore operate at complementary levels, network slicing provides broader resource isolation, whereas \ac{qos} Flow-level policy control enables dynamic priority and preemption among concurrent communications.

The management of mission-critical services in \ac{b5g} networks is explored in \cite{orch}, where the authors describe an architectural framework for cloud-native mission-critical service deployment, network slicing, edge placement, service scaling, migration, and machine learning assisted overload prediction. Evaluated scenarios include scaling resources under increasing service demand, creating dedicated slices for mission-critical applications, and migrating edge-hosted services following failures or \ac{qos} degradation. In contrast, the cross-layer mechanism considered in this work operates at the granularity of an individual active \ac{mcx} communication where changes in operational state can trigger modifications to its network-level priority and preemption treatment without scaling or migrating the service instance or reconstructing the communication session. At the deployment level, this work additionally evaluates the placement of user plane and media functions according to the resulting end-to-end traffic path, particularly whether mission-critical media can remain at the tactical edge while signaling and centralized control traverse a constrained \ac{ntn} backhaul.

A related closed-loop approach is proposed in \cite{nwdaf}, where an \ac{mcx} server, \ac{nwdaf}, telemetry module, and orchestrator cooperate to detect and predict overload conditions. \ac{mcx}-level metrics such as the number of registered users and active private or group calls are used to support proactive scaling decisions. This demonstrates the value of incorporating mission-critical application state into network-management decisions. The resulting control loop, however, primarily operates at the service-orchestration level, adapting the resources assigned to service instances. The problem considered in this work operates at a different granularity, in our case, operational changes within an active \ac{mcx} communication are used to trigger policy modifications affecting the treatment of its associated 5G \ac{qos} Flow.

Resilience evaluation is addressed in \cite{resilience_mcc}, where the authors propose a framework for measuring 5G resilience in the context of mission-critical communications. Their approach identifies \ac{3gpp}-oriented \acp{kpi}, relates communication-system degradation to operational mission impact, and evaluates resilience improvements enabled by multi-access mechanisms such as \ac{atsss} and \ac{mptcp}. The work is particularly relevant because it treats resilience as a dynamic process involving degradation, adaptation, and recovery rather than evaluating only steady-state throughput or latency. Such an approach is well suited to public-safety systems that must continue providing critical functions while network conditions evolve during an incident.

Overall, these studies demonstrate that future mission-critical platforms will increasingly require the Application Server to operate in coordination with mechanisms across the underlying network infrastructure. Network slicing, edge placement, telemetry, analytics, orchestration, and resilience mechanisms provide a foundation for adapting both service and network behavior to changing operational and network conditions.

\subsection{UAV, NTN, and tactical edge 3D Public-Safety Networks}

Deployable communication systems represent an important direction for public-safety operations because emergency incidents frequently occur where fixed infrastructure is unavailable, damaged, congested, or insufficient. Portable radio systems, mobile base stations, aerial platforms, and non-terrestrial backhaul can be used to rapidly establish temporary coverage and restore communication capabilities. The overview in \cite{deployable_5g} examines temporary 5G networks designed to provide coverage, capacity, or redundancy when conventional infrastructure cannot satisfy operational requirements. The authors discuss deployable base stations, cells-on-wheels, aerial platforms, alternative architectural configurations, coexistence and performance constraints. This work establishes the broader architectural motivation for portable private 5G systems in public-safety scenarios and highlights the need to jointly consider radio access, backhaul, mobility, and service availability.

\ac{uav}-assisted communication has been widely investigated as an enabler for rapidly deployable B5G and 6G networks, with aerial platforms acting as base stations, relays, backhaul nodes, and mobile edge-computing platforms~\cite{li_uav5g,uavsurvey,uavsurvey2}. The literature demonstrates their potential for restoring or extending coverage while identifying persistent challenges involving interference, mobility, energy, resource allocation, and backhaul reliability. However, these studies predominantly address the communication and computing infrastructure independently of the mission-critical service layer.

Khan et al. \cite{mcvideouav} consider a \ac{uav}-assisted architecture for maintaining public-safety video services under degraded terrestrial coverage. The results demonstrate the potential of aerial relays and adaptive radio-resource allocation for emergency video delivery. However, video is modeled as a generic streaming service rather than as a complete \ac{mcvideo} session, and the work does not consider \ac{mcx} group management, service level priority, signaling procedures, or coordination with other mission-critical services. It therefore demonstrates the value of aerial video connectivity without addressing how standardized mission-critical service semantics interact with the underlying deployable network.

The integration of terrestrial and non-terrestrial networks for mission-critical services is studied in \cite{ntn}. The authors analyze TN–NTN architectural alternatives for \ac{mcx} support and discuss mechanisms for vertical mobility between terrestrial and satellite domains. Such integration is highly relevant in disasters, maritime operations, and remote areas where terrestrial infrastructure may be absent or severely impaired. At the same time, \ac{ntn} connectivity introduces characteristics such as increased propagation delay, variable capacity, intermittent availability, and heterogeneous mobility behavior that may affect signaling, media transport, and service continuity. These constraints motivate architectures in which critical \ac{mcx} functions can operate locally and adapt their behavior to the availability and quality of external backhaul. However, architectural support for TN–NTN integration does not by itself determine which \ac{mcx} control and media functions should remain local, nor quantify the service level impact of moving latency-sensitive media traffic across the non-terrestrial segment. This placement question is therefore explicitly evaluated in this work.

Moving from architectural studies toward experimental systems, Figaro et al. \cite{leouav} demonstrate the feasibility of mounting a \ac{leo} satellite terminal on a \ac{uav} to restore backhaul connectivity in \ac{ppdr} scenarios. Their prototype provides local Wi-Fi access and experimentally characterizes the performance of the aerial satellite link, showing that UAV-supported \ac{ntn} connectivity can provide practical emergency backhaul. The system, however, does not incorporate 5G access or standardized mission-critical services.

Extending this concept toward a more complete mobile-network architecture, Correia et al. \cite{uavntns} previously developed within our research group a deployable \ac{ppdr} platform combining multi-\ac{uav} networking, aerial 5G access, edge user plane processing, and \ac{leo} satellite backhaul. The experimental results demonstrated the feasibility of integrating these components within a rapidly deployable emergency communication infrastructure. The present work builds upon this infrastructure and extends it from connectivity and network-service provision toward an end-to-end mission-critical service platform. In particular, standardized \ac{mcptt}, \ac{mcvideo}, and \ac{mcdata} functionality is integrated with \ac{ims}, application-driven 5G policy control, and edge media processing. This enables the underlying aerial 5G–\ac{ntn} infrastructure to be evaluated from the perspective of mission-critical service behavior, including dynamic priority transitions and the placement of \ac{mcx} media functions relative to the non-terrestrial backhaul.

These studies demonstrate how terrestrial, aerial, and satellite components can be combined into emerging 3D public-safety networks. However, most \ac{uav}-\ac{ntn} oriented research treats emergency applications as generic traffic sources and does not exploit the operational semantics available within standardized \ac{mcx} services. Consequently, limited experimental evidence exists on how \ac{mcx} signaling, group media, floor control, and application priority behave when control and media functions are distributed between the edge and remote infrastructure connected through \ac{ntn}. In particular, the trade-off between centralized media anchoring and local edge media processing remains important when the external backhaul exhibits increased latency or constrained capacity.

The reviewed literature reveals complementary advances across mission-critical service implementation, programmable 5G networking, and deployable aerial and non-terrestrial infrastructures. However, these capabilities are generally investigated independently or in subsets. Table~\ref{tab:related_work_comparison} compares representative works according to the capabilities most directly related to the research questions addressed in this paper. The comparison emphasizes multi-service \ac{mcx} operation, application-aware network priority, dynamic interaction between \ac{mcx} state and 5G policy control, evaluation under resource contention, edge function placement and non-terrestrial connectivity.

\begin{table*}[t]
\centering
\caption{Comparison of representative approaches related to the proposed MCX platform.}
\label{tab:related_work_comparison}

\renewcommand{\arraystretch}{1.20}
\setlength{\tabcolsep}{5pt}

\resizebox{\textwidth}{!}{%
\begin{tabular}{lccccccc}
\toprule

\textbf{Reference} &
\shortstack[c]{\textbf{MCPTT +}\\
               \textbf{MCVideo +}\\
               \textbf{MCData}} &
\shortstack[c]{\textbf{Experimental}\\
               \textbf{5G}} &
\shortstack[c]{\textbf{MCX-aware}\\
               \textbf{network QoS}} &
\shortstack[c]{\textbf{Dynamic MCX}\\
               \textbf{state $\rightarrow$}\\
               \textbf{5GC policy}} &
\shortstack[c]{\textbf{5G priority /}\\
               \textbf{preemption}\\
               \textbf{under contention}} &
\shortstack[c]{\textbf{MCX edge /}\\
               \textbf{function}\\
               \textbf{placement}} &
\shortstack[c]{\textbf{NTN}\\
               \textbf{integration}} \\
\midrule

\cite{lteptt}
& & & \(\circ\) & & \(\circ\) & & \\

\cite{mcptt_5gnsa}
& & \(\circ\) & \(\circ\) & & & \checkmark & \\

\cite{ran_limitations}
& & & \(\circ\) & & & & \\

\cite{open_mcptt}
& & \checkmark & \(\circ\) & & & & \\

\cite{frmcs}
& \(\circ\) & \checkmark & \checkmark & & & \(\circ\) & \\

\cite{slicing}
& \(\circ\) & \checkmark & \checkmark & & \(\circ\) & \(\circ\) & \\

\cite{orch}
& \(\circ\) & \(\circ\) & \(\circ\) & \(\circ\) & \(\circ\) & \checkmark & \\

\cite{nwdaf}
& \(\circ\) & \(\circ\) & \(\circ\) & \(\circ\) & & \(\circ\) & \\

\cite{ntn}
& \(\circ\) & \(\circ\) & \(\circ\) & & & \(\circ\) & \checkmark \\

\cite{leouav}
& & & & & & & \checkmark \\

\cite{uavntns}
& & \checkmark & & & & & \checkmark \\

\midrule

\textbf{This work}
& \checkmark
& \checkmark
& \checkmark
& \checkmark
& \checkmark
& \checkmark
& \checkmark \\

\hline
\end{tabular}%
}

\vspace{1mm}

\footnotesize{
\checkmark: implemented or explicitly evaluated as a central mechanism;
\(\circ\): architecturally considered or indirectly evaluated without
end-to-end implementation of the stated capability;
blank: not addressed.
}

\end{table*}

\section{Use Cases and System Requirements}
\label{sec:use_cases}

\begin{figure}
	\centering
	\includegraphics[width=1\columnwidth]{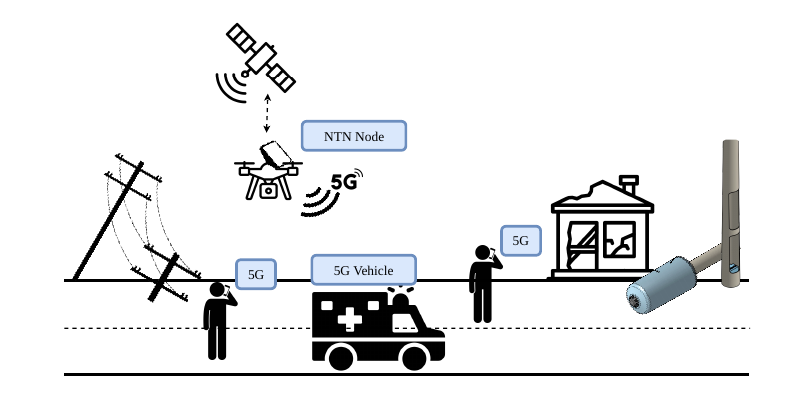}
	\caption{Disaster affected area supported by a private 5G network with aerial nodes to extend coverage via \ac{ntn}.}
	\label{fig:use_case}
\end{figure}

The proposed platform targets public-safety operations over constrained, deployable 3D communication infrastructures spanning terrestrial, aerial, and non-terrestrial network segments.
A disaster affects an area where terrestrial communication infrastructure is unavailable or partially degraded. A deployable private 5G network provides local access, while aerial nodes extend coverage and an \ac{ntn} link provides external backhaul, as demonstrated in Figure~\ref{fig:use_case}. Responders require group voice, location, operational data, and selected video streams. 

A large-scale incident may involve multiple responder teams and operational groups sharing the same radio and network resources while background traffic is also present. Within an individual \ac{mcptt} group, application-level floor control determines which affiliated member is authorized to transmit and allows a higher-priority participant to preempt the current floor holder. However, this mechanism does not establish a resource priority hierarchy across independent communications belonging to different groups. Under network congestion, flows associated with different groups may therefore compete for the same constrained resources despite having different operational importance. Moreover, this importance may change dynamically, for example when a participant in one group enters an emergency state. This scenario motivates mapping \ac{mcx} operational context to network-level \ac{qos} and \ac{arp}, allowing priority and preemption to be enforced across competing communications independently of their group membership.


The two scenarios lead to a set of requirements that directly influence the proposed architecture.

\begin{itemize}
    \item \textit{R1: Integrated multi-service MCX operation.}
The platform shall support concurrent \ac{mcptt}, \ac{mcvideo}, and
\ac{mcdata} operation together with the common user, group, location, and
emergency context required to coordinate these services.
    \item \textit{R2: Cross-layer priority and resource control.}
The platform shall distinguish application-level priority from network-level resource priority. Within an \ac{mcptt} group, the configured user priority and dynamic emergency state shall determine floor access and application-level preemption according to the service policy. At the network layer, the service type shall determine the baseline \ac{5qi}, while the configured user priority and emergency state shall determine the applicable \ac{arp} policy. This mapping shall enable differentiated admission, retention, and preemption treatment among competing \ac{qos} Flows, including communications associated with different groups. Priority changes occurring during an active communication shall be applicable without reconstructing the corresponding service session or changing its service-specific \ac{5qi}.
    \item \textit{R3: Edge-deployment.}
Latency-sensitive and operationally critical \ac{mcx} functions shall be
deployable within the tactical edge network so that local communication can
continue with reduced dependence on external backhaul.
    \item \textit{R4: Heterogeneous 3D connectivity and backhaul.}
The platform shall support deployable private 5G operation across terrestrial, aerial, and non-terrestrial network segments, forming a heterogeneous 3D public-safety network, and accommodate variations in backhaul delay, capacity, and availability.
\end{itemize}

Table~\ref{tab:requirements_mapping} summarizes how the requirements are reflected in the implemented platform.

\begin{table*}[t]
\centering
\caption{Traceability between system requirements, platform mechanisms,
and experimental validation.}
\label{tab:requirements_mapping}
\begin{tabularx}{\textwidth}{@{}p{0.06\textwidth}p{0.20\textwidth}X p{0.23\textwidth}@{}}
\toprule
\textbf{Req.} &
\textbf{Requirement} &
\textbf{Platform mechanism} &
\textbf{Experimental validation} \\
\midrule

R1 &
Integrated multi-service MCX &
MCPTT, MCVideo, and MCData services sharing common user, group, location,
and emergency context &
Functional validation \\

R2 &
Cross-layer priority and resource control &
Service type and operational priority translated through the \ac{sip} \ac{as} and N5 into \ac{5qi} and \ac{arp} treatment &
5QI differentiation, ARP admission/preemption, dynamic emergency priority,
and reaction time \\

R3 &
Edge-local deployment &
Edge-deployable UPF, RTPengine, and MCX components &
Central/edge placement\\

R4 &
Heterogeneous access and backhaul &
Private 5G access, aerial deployment, and NTN backhaul &
Terrestrial/NTN comparison\\

\bottomrule
\end{tabularx}
\end{table*}

\section{Cross-layer QPP Model for MCX}
\label{sec:cross_layer_priority}

The proposed model distinguishes between the communication requirements of an \ac{mcx} service and the operational priority associated with the users participating in it. The former determines the baseline network treatment required by the service type, whereas the latter is derived from application-level information, including the configured user priority and the current emergency state. Operational priority determines the relative importance of a communication and is used to provide preferential treatment when multiple users or traffic flows compete for constrained resources.

In the 5G system, service differentiation is applied through \ac{qos} Flows identified by a \ac{qfi}. The \ac{5qi} associated with each flow defines its baseline forwarding characteristics. In the considered platform, mission-critical signaling, \ac{mcptt}, \ac{mcvideo}, and \ac{mcdata} are therefore associated with service-specific \ac{5qi} profiles, as summarized in Table~\ref{tab:mcx_5qi} according to the standardized \ac{5qi} characteristics defined in \ac{3gpp} TS~23.501. These profiles remain associated with the service independently of the operational priority of the user generating the traffic.

\begin{table}[t]
\centering
\caption{MCX-relevant standardized 5QI values.}
\label{tab:mcx_5qi}
\begin{tabular*}{\columnwidth}{@{\extracolsep{\fill}}lcccc@{}}
\toprule
\textbf{Service} & \textbf{5QI} & \textbf{Type} & \textbf{Priority} & \textbf{PDB} \\
\midrule
MC delay-sensitive signaling & 69 & Non-GBR & 5 & 60 ms \\
MCPTT voice & 65 & GBR & 7 & 75 ms \\
MCVideo & 67 & GBR & 15 & 100 ms \\
MCData & 70 & Non-GBR & 55 & 200 ms \\
\bottomrule
\end{tabular*}
\end{table}

Operational priority is maintained by the \ac{mcx} service layer. Each authenticated user is assigned a configured priority as part of the user profile, which establishes the user’s baseline application-level priority. This priority may change dynamically according to the operational state, most notably when the user activates an emergency condition. The resulting operational priority can then be applied at different layers of the system: within an \ac{mcptt} group, it influences floor control decisions, while at the network layer, it can be translated into differentiated admission, retention, and preemption treatment.

\subsection{Application-Level and Network-Level Preemption}

Priority is enforced at two complementary levels. At the application layer, \ac{mcptt} floor control resolves competition between participants affiliated with the same group. When multiple users request permission to transmit, their configured priority and emergency state can be used by the \ac{sip} \ac{as} to determine floor ownership. A higher-priority request may therefore revoke the floor from the current participant and assign it to another member of the same group.

Application-level floor preemption does not, however, resolve competition between independent groups. Consider two simultaneously active groups $G_A$ and $G_B$. Floor control can determine which participant transmits within $G_A$ and which participant transmits within $G_B$, but it does not determine how the corresponding media flows compete for shared 5G resources. Consequently, an emergency communication in $G_A$ may still compete at the network layer with routine communication from $G_B$.

The proposed mechanism addresses this second level of contention by translating \ac{mcx} operational priority into network-level \ac{arp} treatment. Whereas floor preemption establishes a priority hierarchy among participants of the same group, \ac{arp} provides admission, retention, and preemption differentiation among competing \ac{qos} Flows independently of their group membership. The two mechanisms, therefore, provide complementary priority enforcement at the application and network layers.

\subsection{Cross-Layer Priority Mapping}

The proposed model distinguishes between a user’s configured priority profile and its dynamic operational state. The network-level priority policy is represented by the three components of \ac{arp},
\begin{equation}
P_i =
\left(
PL_i,PC_i,PV_i
\right),
\end{equation}

where $PL_i$ denotes the \ac{arp} priority level, $PC_i$ the preemption capability, and $PV_i$ the preemption vulnerability.

For implementation, five application-level priority classes are defined, with lower values representing higher operational priority. Levels~2–5 establish the configured hierarchy between users during normal operation, and are used by the \ac{mcx} service layer for application-level decisions, including floor arbitration. Level~1 represents the highest application priority and is reserved for the emergency state. Each application priority level is additionally mapped to a corresponding network-level \ac{arp} policy. The mapping used in the experimental platform is summarized in Table~\ref{tab:priority_mapping}.

\begin{table}[t]
\centering
\caption{Proposed Application and network priority mapping.}
\label{tab:priority_mapping}
\begin{tabular*}{\columnwidth}{@{\extracolsep{\fill}}lcccc@{}}
\toprule
\textbf{Operational class} & \textbf{App. PL} & \textbf{ARP PL} & \textbf{PC} & \textbf{PV} \\
\midrule
Emergency          & 1 & 1 & Yes & No \\
Incident commander & 2 & 3 & Yes  & Yes \\
Team leader        & 3 & 5 & No  & Yes \\
Responder           & 4 & 7 & No  & Yes \\
Low priority        & 5 & 9 & No  & Yes \\
\bottomrule
\end{tabular*}
\end{table}

We refer to the combination of service class and operational resource policy as \acf{qpp}. Quality is represented by the service-specific \ac{5qi}; Priority by the \ac{arp} level $PL_i$; and preemption by $PC_i$ and $PV_i$. The service type determines the \ac{5qi}, while the configured user priority and emergency state select $P_i$. An emergency transition can therefore update $P_i$ on an existing flow while preserving its \ac{5qi} and QFI.

Application priority level~1 is reached dynamically rather than assigned as the user’s normal configured priority. When an authorized user activates emergency mode, both its application-level priority and the corresponding network-level priority policy are temporarily elevated. For example, a responder operating with application priority level~4 and the baseline network policy
\begin{equation}
P_i^{\mathrm{base}} =
(7,\mathrm{No},\mathrm{Yes})
\end{equation}

transitions to application priority level~1 and the emergency network policy
\begin{equation}
P_i^{\mathrm{emergency}} =
(1,\mathrm{Yes},\mathrm{No}).
\end{equation}

At the application layer, this transition gives the emergency user preferential treatment during floor arbitration, and enables floor preemption according to the configured service policy. At the network layer, the corresponding \ac{arp} transition provides preferential admission and retention treatment, and enables preemption of vulnerable lower-priority resources. Importantly, the transition modifies operational priority rather than the service class; the service-specific \ac{5qi} remains unchanged. Once the emergency state is cleared, both the application priority and the network-level \ac{arp} policy return to the baseline values associated with the user’s configured priority.

\subsection{Dynamic QPP Enforcement}

To enforce the priority transition described above on an already established communication, the updated operational state is propagated from the \ac{mcx} user to the \ac{sip} \ac{as}. The Application Server then coordinates the transition at both the application and network layers. At the application layer, it updates the user state maintained by the \ac{cms}, allowing the new operational priority to be considered in subsequent service level decisions, including \ac{mcptt} floor arbitration and preemption. At the network layer, the Application Server acts as a trusted \ac{af} and requests the corresponding policy modification from the \ac{pcf} through N5. The same operational transition is therefore reflected in both the application-level priority used by the \ac{mcx} service logic, and the network level priority applied to the associated \ac{qos} Flow.

To support this behavior, the Open5GS \ac{pcf} and \ac{smf} were extended to dynamically modify the \ac{arp} associated with an existing \ac{qos} Flow through its N5 AppSession. The initial implementation was contributed to the Open5GS project, and subsequently incorporated upstream through Pull Request~\#4794\footnote{\url{https://github.com/open5gs/open5gs/pull/4794}}. The resulting implementation maps the N5 \texttt{resPrio} attribute to an operator-configurable \ac{arp} priority level, and independently applies \texttt{preemptCap} and \texttt{preemptVuln}. ARP-only updates are propagated through the existing \ac{qos} Flow modification procedures while preserving the established \ac{qfi}, \ac{5qi}, and negotiated bit-rate parameters.

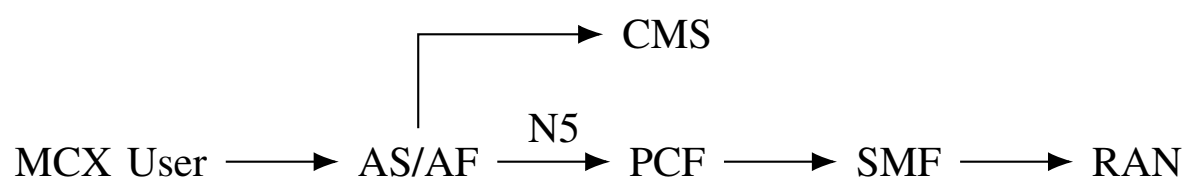
\begin{figure}
\centering
\begin{tikzpicture}[
    node distance=0.4cm and 0.75cm,
    every node/.style={font=\small},
    >=Latex
]

\node (mcx) {MCX User};
\node (as)  [right=of mcx] {AS/AF};
\node (pcf) [right=of as] {PCF};
\node (smf) [right=of pcf] {SMF};
\node (ran) [right=of smf] {RAN};
\node (cms) [above=of pcf] {CMS};

\draw[->] (mcx) -- (as);
\draw[->] (as) -- node[midway, above]{N5} (pcf);
\draw[->] (pcf) -- (smf);
\draw[->] (smf) -- (ran);

\draw[->] (as) |- (cms);

\end{tikzpicture}
\caption{Cross-layer propagation of an MCX operational priority
transition to the application and network layers.}
\label{fig:priority_control_path}
\end{figure}

The resulting cross-layer control path is illustrated in Figure~\ref{fig:priority_control_path}. Upon an operational priority transition, the Application Server updates the corresponding user state in the \ac{cms} while, in parallel, requesting the network-policy modification through N5.

Emergency deactivation follows the same control path in the opposite policy direction. The \ac{cms} restores the configured application-level priority associated with the user profile, while the N5 policy update restores the corresponding baseline \ac{arp} configuration of the active \ac{qos} Flow. This coordinated mechanism ensures that a change in operational priority is consistently reflected at both layers. Within an \ac{mcptt} group, the updated application-level priority affects floor arbitration and enables application-level preemption according to the service policy. Across independently active communications, the corresponding \ac{arp} update establishes differentiated admission, retention, and network level preemption treatment. Communications using the same service class, including simultaneous \ac{mcptt} sessions belonging to different groups, can therefore retain the same standardized \ac{5qi} while receiving different treatment according to their current operational priority.

An \ac{arp} update should be distinguished from a packet forwarding priority update. The proposed transition changes the priority and preemption attributes of a given \ac{qos} flow; however, it will not affect the \ac{5qi}, packet delay budget, guaranteed bit rate, or the application layer media policy for the flow. Acceptance of an N5 request by the \ac{pcf} confirms that the request was accepted, but not that the resulting policy has been enforced by the \ac{ran}. Confirmation at the \ac{ran} provides an additional control plane indication. However, neither indication by itself would prove preferential packet scheduling, lower media latency, or preemptive treatment of another flow. These are items that can be tested through experimentation when resources are being utilized with measurements taken on the impacted flows. Group floor arbitration currently utilizes a custom \ac{sip} based procedure as defined in Section~\ref{sec:architecture}.\ref{subsec:ims_sip_as}; its behavior should therefore be distinguished from standardized \ac{mcptt} floor control, and interoperability should not be assumed.

\section{Proposed Architecture}
\label{sec:architecture}

This section presents the proposed \ac{qpp}-aware \ac{mcx} architecture for a tactical edge 3D public-safety network. The architecture integrates mission-critical service logic, \ac{ims} signaling, media processing, 5G policy control, terrestrial and aerial access, tactical edge processing, and non-terrestrial backhaul.

\subsection{System Overview}

The platform follows a layered architecture comprising an \ac{mcx} User Equipment Layer, an \ac{mcx} Service Layer, and a Network Infrastructure Layer. This separation decouples user applications from mission-critical service logic and from the underlying 5G transport network, allowing each layer to evolve independently while remaining coordinated through common service state, signaling, media control, and network-level \ac{qos} mechanisms.

A central design principle is the separation between control and user plane functions. Core control functions, including the \ac{amf}, \ac{smf}, \ac{pcf}, \ac{udm}, and \ac{ausf} can remain at a central or tactical command location, while the \ac{upf} is deployable close to the \ac{ran}. This enables local breakout toward edge-hosted service and media functions and reduces the dependence of mission-critical user traffic on constrained or intermittent backhaul. The distinction is particularly relevant when \ac{ntn} provides connectivity between the tactical network and central service infrastructure, where control signaling may traverse the non-terrestrial link while locally anchored user plane traffic can remain within the incident network.

The media plane is decoupled from the signaling and service-control functions. The \ac{sip} \ac{as} maintains session and group state, validates membership, manages floor control, and coordinates media session negotiation through \ac{sdp}, while RTPengine provides the dedicated \ac{rtp} media-relay function. This separation allows media processing to be deployed independently of the \ac{as} and positioned closer to the \ac{upf} and \ac{ran} when required. In tactical deployments, RTPengine can therefore be placed at the edge so that local mission-critical media remains within the incident network even when signaling and central service functions are reached through the \ac{ntn} backhaul.

The current implementation uses unicast media distribution with persistent media legs between RTPengine and affiliated participants. This allows explicit per-participant media state to be maintained, while the forwarding graph is dynamically modified according to floor ownership, without reconstructing the established media sessions. The resulting architecture provides both media control and a controlled baseline for evaluating the scalability and resource cost of unicast group communication.

\begin{figure}[H]
	\centering
	\includegraphics[width=\columnwidth]{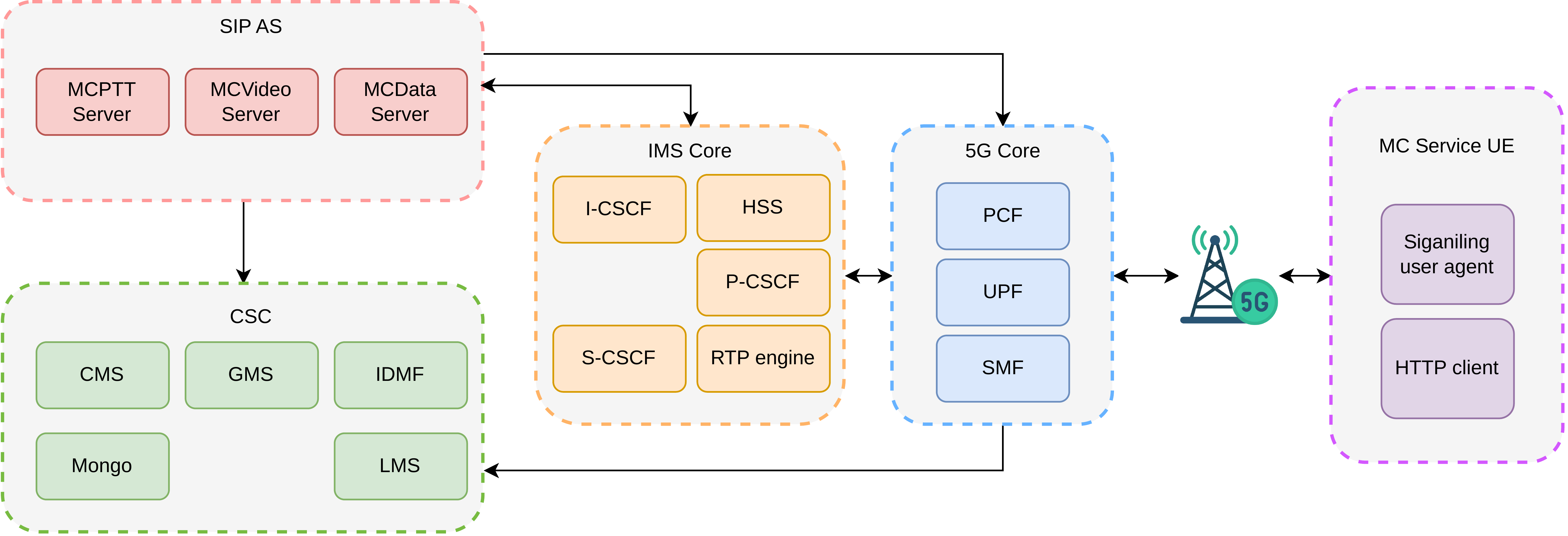}
	\caption{Logical and deployment architecture of the proposed QPP-aware MCX platform across the tactical edge 3D network.}
	\label{fig:architecture}
\end{figure}

The implemented platform integrates the Android responder application, dispatcher, common \ac{mcx} services, Kamailio-based \ac{ims}, the custom \ac{sip} Application Server, RTPengine, and an Open5GS-based private 5G \ac{sa} network. Its modular design supports different deployment configurations, including disaggregated \ac{upf} and media-relay placement, aerial \ac{gnb} deployment and \ac{ntn} backhaul. The complete logical architecture is shown in Figure~\ref{fig:architecture}.

\subsection{MCX User Equipment Layer and Dispatcher}
\label{subsec:mcx_ue_layer}

The \ac{mcx} User Equipment Layer comprises the responder application, dispatcher interface, and specialized endpoints associated with \acp{uav} and \acp{ugv}. These endpoints provide access to the mission-critical services and expose operational context to the service layer, including user identity, group affiliation, configured priority, emergency state, location, and connectivity status.

The responder endpoint is an Android application supporting authentication, \ac{sip} registration, group affiliation, \ac{mcptt} communication, \ac{mcvideo}, direct and group \ac{mcdata}, location sharing, and emergency activation. Voice and video are encoded using Opus and H.264, respectively, and transported using \ac{rtp}/UDP. Signaling and media state are maintained separately, allowing service control and media transport to evolve independently. Offline map support further reduces dependence on external Internet connectivity during degraded-infrastructure operation.



The dispatcher provides a browser-based operational interface for incident, user, and group management and for real-time situational awareness. It exposes responder and tactical-asset locations, connectivity and emergency states, mission-critical video streams, and operational group information. Current and historical location data can additionally be used to reconstruct mobility during experiments and post incident analysis, while live video streams are exposed through a low-latency WebRTC interface.




\acp{uav} and \acp{ugv} can participate as specialized \ac{mcx} endpoints or gateways, publishing location, video, telemetry, and connectivity information and being associated with operational groups and incidents through the same common service functions used by human operated endpoints. This allows tactical assets to participate in the mission-critical service environment rather than being treated as independent external sensors. Their communications can consequently be associated with the same service-specific \ac{qos} and operational-priority mechanisms applied to other \ac{mcx} endpoints.

\begin{figure}
	\centering
	\includegraphics[width=1\columnwidth]{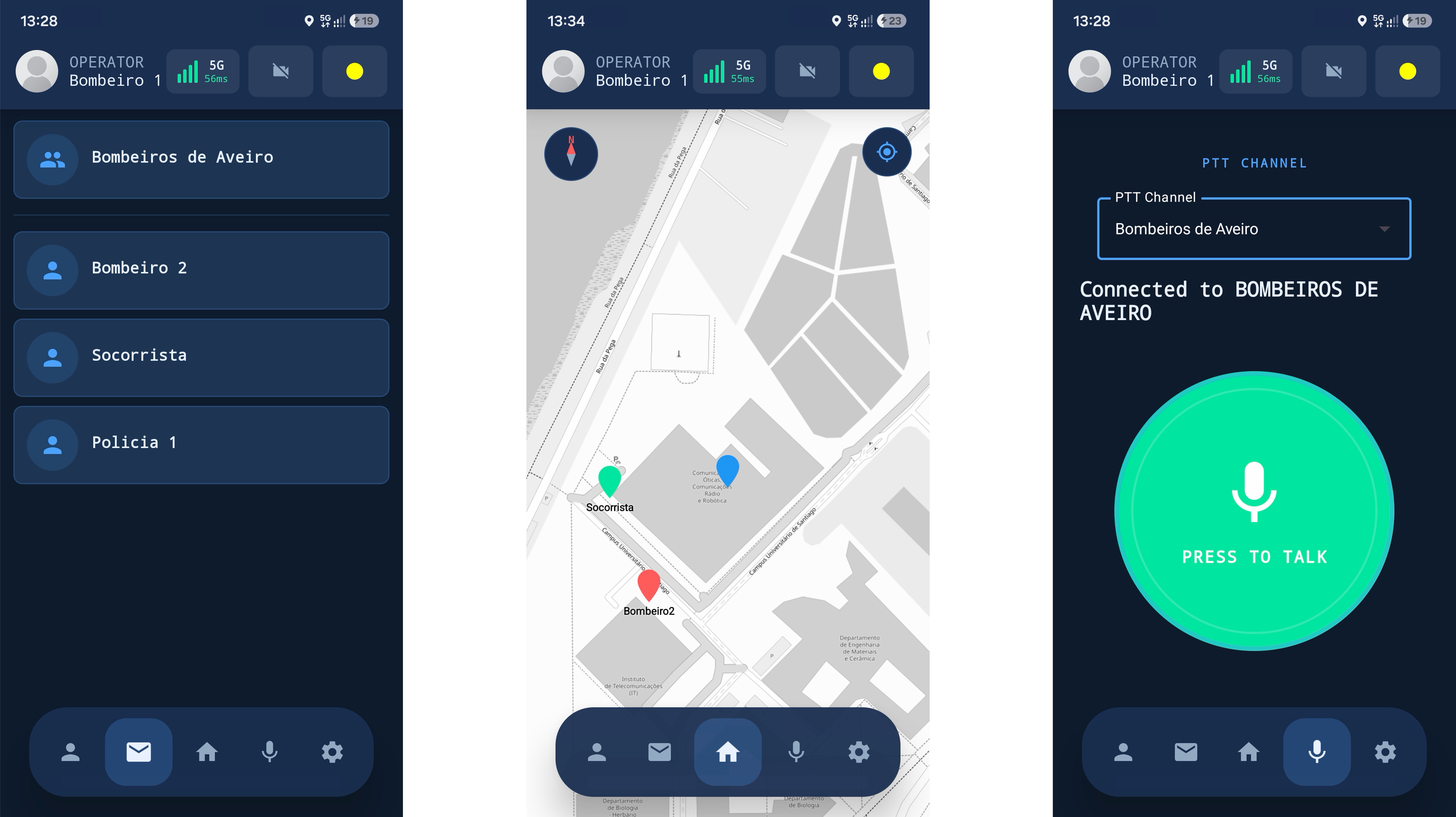}
	\caption{Android MCX Application screens, from left to right: Private and group messages; map with the location of each group's member, and their current state in the application; and push to talk feature inside a group.}
	\label{fig:application}
\end{figure}

\begin{figure}
	\centering
	\includegraphics[width=1\columnwidth]{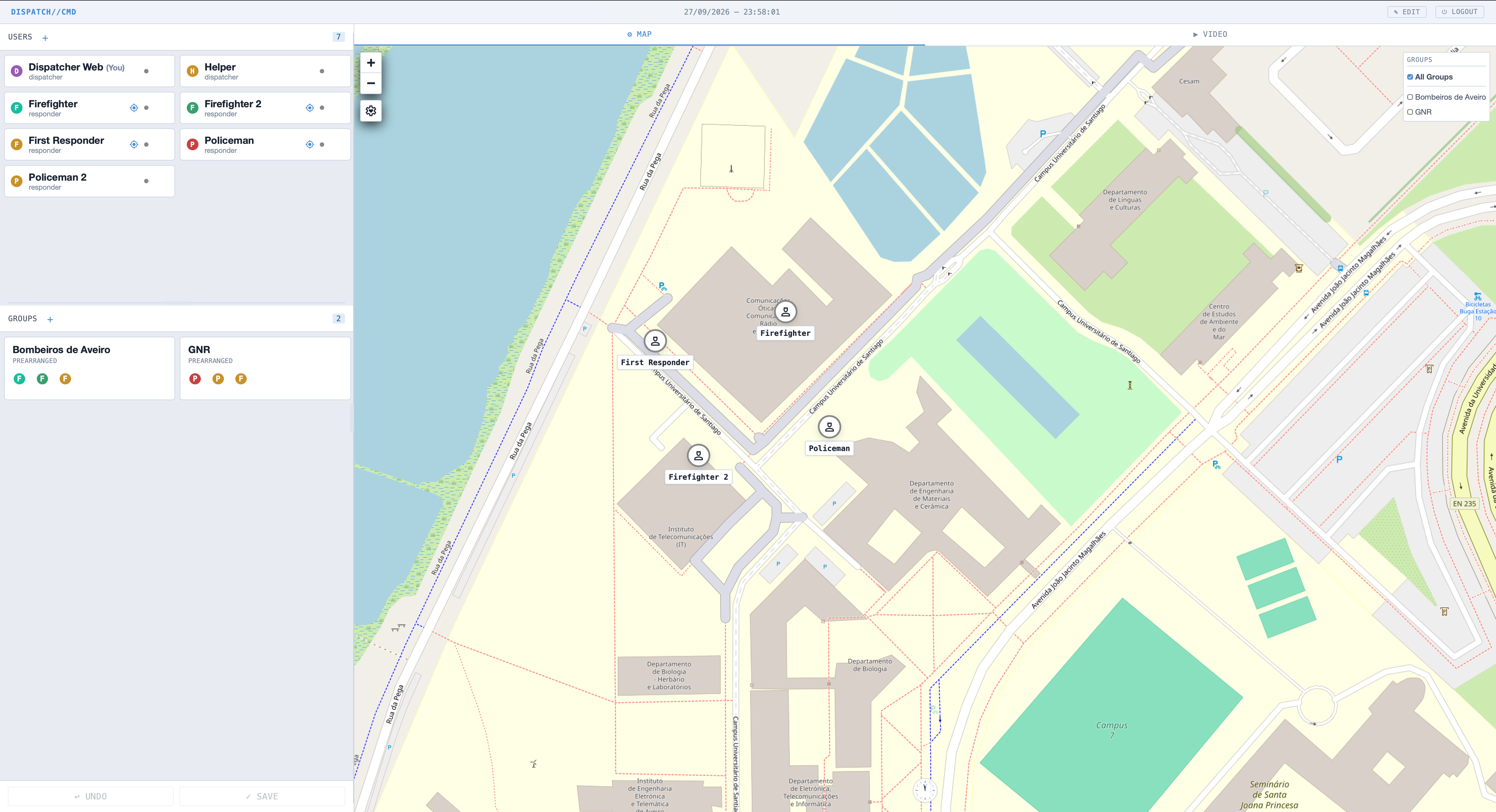}
	\caption{Dispatcher MCX screen: manages the existent groups and members, their priorities and states, and the position of each member in the map.}
	\label{fig:dispatcher}
\end{figure}

Figures~\ref{fig:application} and~\ref{fig:dispatcher} show the responder and dispatcher interfaces, respectively.

\subsection{MCX Common Services Core}
\label{subsec:mcx_csc_layer}

The \ac{mcx} \ac{csc} provides the shared service state required by the responder clients, dispatcher, and service-specific Application Server functions. It separates identity, operational profiles, group membership, and location information from the session-control logic implemented in the \ac{sip} \ac{as}. This follows the general \ac{3gpp} \ac{mcx} principle of sharing common capabilities across multiple mission-critical services while maintaining service-specific signaling and media functions.

The \ac{idmf} provides authentication and maintains the association between authenticated users and their \ac{mcx} identities. Authentication is kept separate from the operational profiles and priority state used by the mission-critical service logic.

The \ac{cms} maintains the operational configuration and user state shared by the \ac{mcx} services. It stores user profiles, roles, configured operational priority, status, and the network identifiers required to associate an \ac{mcx} user with the corresponding 5G communication context. The \ac{sip} \ac{as} uses this information to support \ac{mcptt} floor arbitration and determine the network-priority policy requested through N5. During emergency activation and deactivation, the \ac{sip} \ac{as} updates the user state maintained by the \ac{cms}, allowing the operational priority changes described in Section~\ref{sec:cross_layer_priority} to inform both application-level decisions and network-policy requests.

The \ac{gms} maintains operational groups and their membership, supporting both predefined and dynamically created groups. Group information is exposed to clients and service functions through backend interfaces, including an XCAP-compatible group representation. Before establishing a group communication, the \ac{sip} \ac{as} queries the \ac{gms} to validate the requested group and the membership of the initiating user, ensuring that session and floor control operations are applied only to affiliated participants.

The \ac{lms} maintains current and historical location information for responders and tactical assets. Authenticated location reports are distributed according to group membership and made available to the responder and dispatcher interfaces, providing a common location context for situational awareness and mobility analysis.

\begin{figure}[t]
    \centering
    \includegraphics[width=.85\columnwidth]{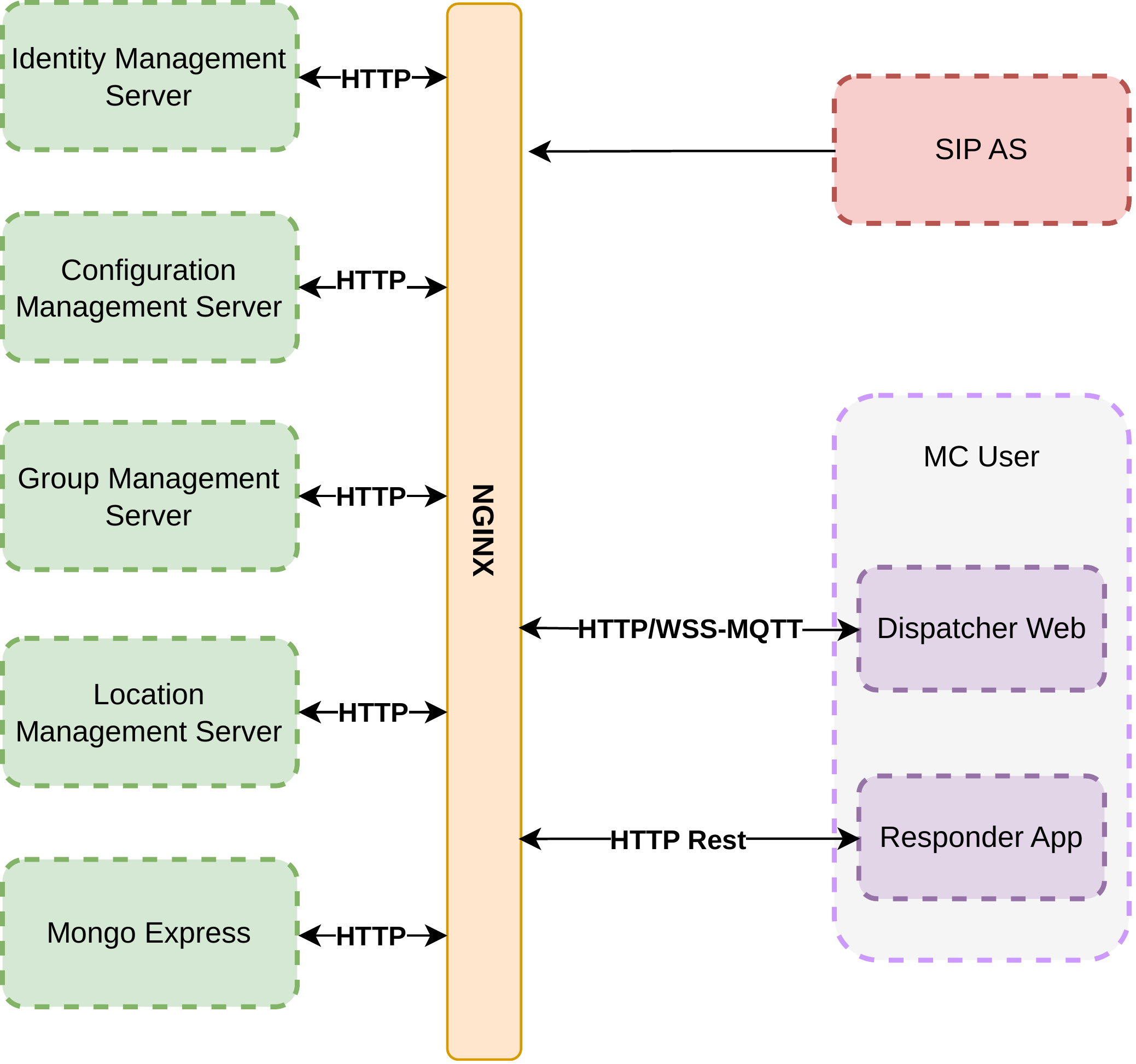}
    \caption{Common Services Core interaction with MCX users, backend services, and the SIP Application Server.}
    \label{fig:csc}
\end{figure}

Together, these functions provide the shared identity, operational, group, and location context used by the service-specific \ac{mcx} functions, while session control, floor arbitration, and network-policy requests remain the responsibility of the \ac{sip} \ac{as}. Figure~\ref{fig:csc} illustrates the interaction between the \ac{mcx} endpoints, the common service functions, and the \ac{sip} \ac{as}.

\subsection{IMS/SIP signaling and Application Server Design}
\label{subsec:ims_sip_as}

Kamailio provides the \ac{ims}/\ac{sip} signaling plane, including \ac{pcscf}, \ac{icscf}, and \ac{scscf} functions for \ac{ue} registration and session routing. Initial Filter Criteria route the relevant mission-critical signaling toward the custom \ac{sip} \acl{as}, which implements the service logic for \ac{mcptt}, \ac{mcvideo}, and \ac{mcdata}.

The \ac{sip} Application Server is organized into service-specific \ac{mcptt}, \ac{mcvideo}, and \ac{mcdata} modules that share the common service functions described in Section~\ref{subsec:mcx_csc_layer} while maintaining independent service state and signaling logic. The \ac{mcptt} module manages group affiliation, floor control, priority, and group-media distribution; the \ac{mcvideo} module manages publishers, listeners, and video forwarding; and the \ac{mcdata} module validates, persists, and distributes direct and group messages.

\subsubsection{MCPTT Group Sessions and Floor Control}

Each \ac{mcptt} group retains its affiliated members, a current floor
holder, and a shared RTPengine call with one persistent bidirectional
media participant and one N5 AppSession per member.
Algorithm~\ref{alg:mcptt-control} describes affiliation
through a \ac{sip} \texttt{INVITE}, and floor control through in-dialog floor messages. The \ac{cms} resolves user identity and
supplies floor priorities, while the \ac{gms} verifies group membership.
Priority comparisons use a lower numeric value for a higher priority. A brief description of the algorithm's operation process is given below:

\begin{itemize}
    \item \underline{Lines 1--9 (affiliation).} The server validates identity
and membership, publishes a stable bidirectional RTPengine participant,
and requires an initial downlink AppSession before accepting the
\texttt{INVITE}. A late member is connected to the current talker
without acquiring the floor.
    \item \underline{Lines 10--16 (arbitration).} A request during a floor
change receives \texttt{486 Busy Here}. When another member is talking,
the server compares their \ac{cms} priorities. The requester preempts only
if its priority is higher, represented by a lower numeric value;
otherwise, the server replies \texttt{486 Busy Here}.
    \item \underline{Lines 17--29 (media switch).} Under the group lock,
the server rechecks floor ownership. If the previous talker has left,
the floor can be granted directly; if another talker has taken over,
the request is rejected. For preemption, the server starts an
asynchronous \texttt{floor-revoked} notification to the former talker,
requests a downlink N5 update, and attempts to block its media.
It then assigns the requester, requests an uplink update, unblocks its
media, and connects it to each listener in turn. A failed RTPengine
operation is logged and does not stop the remaining attempts.
    \item \underline{Lines 30--35 (release).} The request is acknowledged
before the background task requests the downlink policy update, attempts
to block the current talker, and clears the floor. Directional N5
\texttt{PATCH} completion and revocation delivery do not gate a floor
reply. For a grant or preemption, \texttt{200 OK} follows the attempted
RTPengine operations; for release, it precedes them. The established
participants and AppSessions are reused without \ac{sip}/\ac{sdp}
renegotiation. A SIP response alone confirms neither media reception
nor downstream \ac{qos} enforcement.
\end{itemize}

\begin{algorithm}
\caption{MCPTT group affiliation and floor control.}
\label{alg:mcptt-control}
\small
\algrenewcommand\algorithmicfunction{\textbf{On}}
\algrenewcommand\alglinenumber[1]{\footnotesize #1}
\begin{algorithmic}[1]
\Function{GroupInvite}{$u,g$}
    \State Verify $u$ via CMS and membership via GMS
    \State \textbf{if} denied \textbf{then} reject \texttt{INVITE}; \Return
    \State Publish $u$ at RTPengine as bidirectional participant $m$
    \State \textbf{if} publish fails \textbf{then} reject; \Return
    \State \textbf{if} initial DL AppSession POST fails \textbf{then}
    \State \quad Delete $m$; reject \texttt{INVITE} with 503; \Return
    \State Reply \texttt{200 OK} with SDP answer
    \State \textbf{if} talker active \textbf{then} queue connect to $m$
\EndFunction
\Statex
\Function{FloorRequest}{$u,g$}
    \State \textbf{if} $u$ not affiliated \textbf{then} reply 481; \Return
    \State \textbf{if} floor lock held \textbf{then} reply 486; \Return
    \State $t\gets$ current talker
    \State \textbf{if} $t\ne\varnothing$ and $t\ne u$ \textbf{then}
    \State \quad Obtain $p_u,p_t$ from CMS
    \State \quad \textbf{if} $p_u\ge p_t$ \textbf{then} reply 486; \Return
    \State \textbf{with} group floor lock:
    \State \quad $h\gets$ current talker
    \State \quad \textbf{if} $h=u$ \textbf{then} reply 200; \Return
    \State \quad \textbf{if} another talker replaced $t$ \textbf{then} reply 486; \Return
    \State \quad \textbf{if} $h=t$ and $t\ne\varnothing$ \textbf{then}
    \State \qquad Start asynchronous \texttt{floor-revoked} to $t$
    \State \qquad Schedule N5 PATCH: $t\to$ DL
    \State \qquad Attempt RTPengine block for $t$; clear floor
    \State \quad Set talker to $u$; schedule N5 PATCH: $u\to$ UL
    \State \quad Attempt RTPengine unblock for $u$
    \State \quad \textbf{for each} other affiliated member $v$:
    \State \qquad Attempt RTPengine connect from $u$ to $v$
    \State \quad Reply \texttt{200 OK}
\EndFunction
\Statex
\Function{FloorRelease}{$u,g$}
    \State \textbf{if} $u$ not talker \textbf{then} reply 200; \Return
    \State Reply \texttt{200 OK}
    \State In background, under the group floor lock:
    \State \quad Schedule N5 PATCH: current talker $\to$ DL
    \State \quad Attempt RTPengine block; clear the floor
\EndFunction
\end{algorithmic}
\end{algorithm}

\subsubsection{MCVideo and MCData Services}

The \ac{mcvideo} module follows a publisher/listener model in which each stream is associated with a publishing \ac{mcx} user and may be consumed by multiple authorized listeners. The Application Server validates the publisher, maintains the stream state, and coordinates RTPengine operations to distribute the media toward subscribed participants. An additional media leg can be created toward MediaMTX to expose selected streams as low-latency WebRTC feeds to the dispatcher. 

The core signaling logic governing this lifecycle is detailed in Algorithm~\ref{alg:mcvideo-control}. The first \texttt{INVITE} targeting a stream key instantiates the publisher and secures the necessary uplink network resources. Subsequent offerless \texttt{INVITE} requests initialize pending listeners, whose media state and downlink \ac{qos} are strictly established upon receiving the \texttt{ACK}, remaining stable without subsequent dialog renegotiation. Finally, the \texttt{BYE} method dictates the teardown scope: a listener leaving only releases its individual subscription, whereas a \texttt{BYE} from the publisher systematically dismantles the entire stream, gracefully terminating the dialogs of all active listeners.

A brief description of the algorithm's operation process is given below:

\begin{itemize}
    \item \underline{Lines 1--11 (stream initiation).} When an \texttt{INVITE} targets a non-existent stream, the server verifies the publisher's identity through the \ac{cms}. Upon authorization, it publishes the user at RTPengine, instantiates the stream, and replies \texttt{200 OK} with the \ac{sdp} answer. It then schedules an uplink N5 \texttt{POST} to secure network resources and optionally attaches a MediaMTX bridge for external access.
    \item \underline{Lines 12--18 (listener subscription).} If the stream is already active, the request is handled as a subscription. The server rejects \texttt{INVITE}s containing an \ac{sdp} offer, validates the user's identity, requests an RTPengine subscription, and registers the user as pending. It then replies \texttt{200 OK} containing an \ac{sdp} offer.
    \item \underline{Lines 19--23 (listener activation).} When a pending listener replies with an \texttt{ACK} containing the \ac{sdp} answer, the server finalizes the RTPengine media subscription, schedules a downlink N5 \texttt{POST} to establish the listener's \ac{qos}, and marks the user as an active listener.
    \item \underline{Lines 24--38 (stream and subscription teardown).} A \texttt{BYE} request initiates teardown procedures and is immediately acknowledged with a \texttt{200 OK}. If the request comes from a listener, the server simply unsubscribes the user from RTPengine, schedules an N5 \texttt{DELETE}, and removes them from the stream. However, if the \texttt{BYE} comes from the publisher, the server systematically dismantles the entire stream: it disconnects the MediaMTX bridge, terminates all active listener dialogs (including their RTPengine and N5 sessions), removes the publisher's resources, and deletes the stream instance.
\end{itemize}

\begin{algorithm}
\caption{MCVideo stream initiation and subscription.}
\label{alg:mcvideo-control}
\begin{algorithmic}[1]
\small
\State \textbf{On} \textsc{VideoInvite}($u, k$) \Comment{$u$ invites stream $k$}
\State \quad \textbf{if} stream $k$ does not exist \textbf{then} \Comment{Publisher workflow}
\State \quad \quad \textbf{if} $u \neq k$ \textbf{then} reject 403; \textbf{return}
\State \quad \quad Verify $u$ via CMS
\State \quad \quad \textbf{if} denied \textbf{then} reject \textsc{Invite}; \textbf{return}
\State \quad \quad Publish $u$ at RTPengine for stream $k$
\State \quad \quad \textbf{if} publish fails \textbf{then} reject 503; \textbf{return}
\State \quad \quad Create VideoStream $k$ with $u$ as publisher
\State \quad \quad Reply \texttt{200 OK} with SDP answer
\State \quad \quad Schedule N5 POST: $u \to$ UL
\State \quad \quad \textbf{if} MediaMTX enabled \textbf{then} attach bridge to stream $k$
\State \quad \textbf{else} \Comment{Listener workflow}
\State \quad \quad \textbf{if} \textsc{Invite} has SDP \textbf{then} reject 400; \textbf{return}
\State \quad \quad Verify $u$ via CMS
\State \quad \quad \textbf{if} denied \textbf{then} reject \textsc{Invite}; \textbf{return}
\State \quad \quad Request RTPengine subscription for $u$ to stream $k$
\State \quad \quad Register $u$ as pending listener
\State \quad \quad Reply \texttt{200 OK} with offer SDP
\Statex
\State \textbf{On} \textsc{Ack}($u, k$)
\State \quad \textbf{if} $u$ is pending listener \textbf{then}
\State \quad \quad Answer RTPengine subscription with $u$'s answer SDP
\State \quad \quad Schedule N5 POST: $k \to$ DL for $u$
\State \quad \quad Mark $u$ as active listener
\Statex
\State \textbf{On} \textsc{Bye}($u, k$)
\State \quad Reply \texttt{200 OK}
\State \quad \textbf{if} $u$ is publisher \textbf{then}
\State \quad \quad Detach MediaMTX bridge from stream $k$
\State \quad \quad \textbf{for each} listener $v$ in stream $k$:
\State \quad \quad \quad Unsubscribe $v$ at RTPengine
\State \quad \quad \quad Send \textsc{Bye} to $v$
\State \quad \quad \quad Schedule N5 DELETE for $v$
\State \quad \quad Delete publisher session at RTPengine
\State \quad \quad Schedule N5 DELETE for $u$
\State \quad \quad Delete VideoStream $k$
\State \quad \textbf{else if} $u$ is listener \textbf{then}
\State \quad \quad Unsubscribe $u$ at RTPengine
\State \quad \quad Schedule N5 DELETE for $u$
\State \quad \quad Remove $u$ from stream $k$
\end{algorithmic}
\end{algorithm}

The \ac{mcdata} module supports direct and group messaging. The Application Server resolves the sender, validates group membership when required, persists the message, and distributes it to the intended recipient or affiliated group members. Unlike the media-oriented services, \ac{mcdata} does not maintain \ac{rtp} state, allowing data-message handling to remain independent of the media relay.

\subsubsection{Cross-Layer Service Control}

The Application Server also acts as the service level decision point for cross-layer network control. The service type determines the baseline \ac{5qi} associated with the communication, while the configured user priority and dynamic emergency state determine the applicable network-priority policy. When an operational transition requires a change in network treatment, the Application Server acts as a trusted \ac{af} and requests the corresponding policy modification from the \ac{pcf} through N5. This allows the \ac{arp} priority level, preemption capability, and preemption vulnerability of an active communication to be modified while preserving its service-specific \ac{5qi}.

This network-level mechanism complements, rather than replaces, the floor control policy enforced by the \ac{mcptt} module. Floor preemption determines which participant may transmit within a group, whereas \ac{arp}-based control differentiates admission, retention, and preemption among competing \ac{qos} Flows, including communications associated with different operational groups.

\section{Experimental Validation and Results}
\label{sec:experimental_validation}

This section experimentally evaluates the proposed \ac{mcx} platform from both functional and performance perspectives. The evaluation progresses from baseline validation of the individual \ac{mcx} services to increasingly demanding scenarios involving concurrent service operation, resource contention, dynamic mission-critical priority and non-terrestrial backhaul. Particular attention is given to the impact of user plane and media-function placement when the tactical network is connected to the central infrastructure through \ac{ntn}. Measurements are collected across the application, signaling, media, 5G Core, and radio domains to characterize both end-to-end service performance and the contribution of individual system components.

\subsection{Testbed Description}

The experimental platform consists of a tactical radio node and a central service node whose interconnection and function placement are varied according to the evaluated scenario. The tactical node is implemented using a Mini-PC running the Amarisoft 5G \ac{gnb} software and connected to a USRP B210 software-defined radio. The radio operates in NR band~n77 with a 40-MHz channel bandwidth. This platform is designed for integration on the aerial node used in the deployable configuration.

The central node is implemented on a Minisforum platform and hosts the Open5GS 5G Core control functions, the Kamailio-based \ac{ims}, the \ac{mcx} Common Services Core, and the custom \ac{sip} Application Server. Depending on the evaluated deployment, the \ac{upf} and RTPengine media relay are either placed with these central functions or moved to the tactical node. This allows the user plane and media path to be modified without changing the mission-critical service-control functions.

Up to 64 5G \acp{ue} are simulated using an Amarisoft UE SimBox Series to provide a scalable and repeatable multi-user environment. The SimBox simulates the \ac{ue} protocol stack and communicates with the \ac{gnb} over the physical \ac{nr} radio interface, such that the resulting traffic traverses the same \ac{ran} and 5G Core paths as that of a physical 5G terminal. For the scalability experiments, each simulated \ac{ue} executes the same \ac{mcx} signaling and media procedures generated by the responder application, including \ac{sip} registration and session control, floor control exchanges, \ac{rtp} media transmission and reception, and \ac{mcdata} transactions, but without the graphical Android user interface. Consequently, the experiments preserve the application-level communication flows and network interactions of the implemented \ac{mcx} client, while allowing up to 64 concurrent \acp{ue} to be exercised under controlled and repeatable conditions. The same \ac{ue} environment and workload generation are used across the different deployment configurations to enable consistent comparison of signaling, media, radio, and network performance.

\subsection{Deployment Configurations}
\label{subsec:deployment_configurations}

Three deployment configurations are considered to progressively introduce non-terrestrial backhaul and edge-local user plane and media processing. The mission-critical service logic remains hosted on the central node in all three configurations, while the location of the \ac{upf} and RTPengine is varied to determine which portions of the user and media paths must go through the external backhaul. The three defined deployment scenarios can be seen in Table~\ref{tab:placement_scenarios}.

\begin{figure}[H]
	\centering
	\includegraphics[width=1\columnwidth]{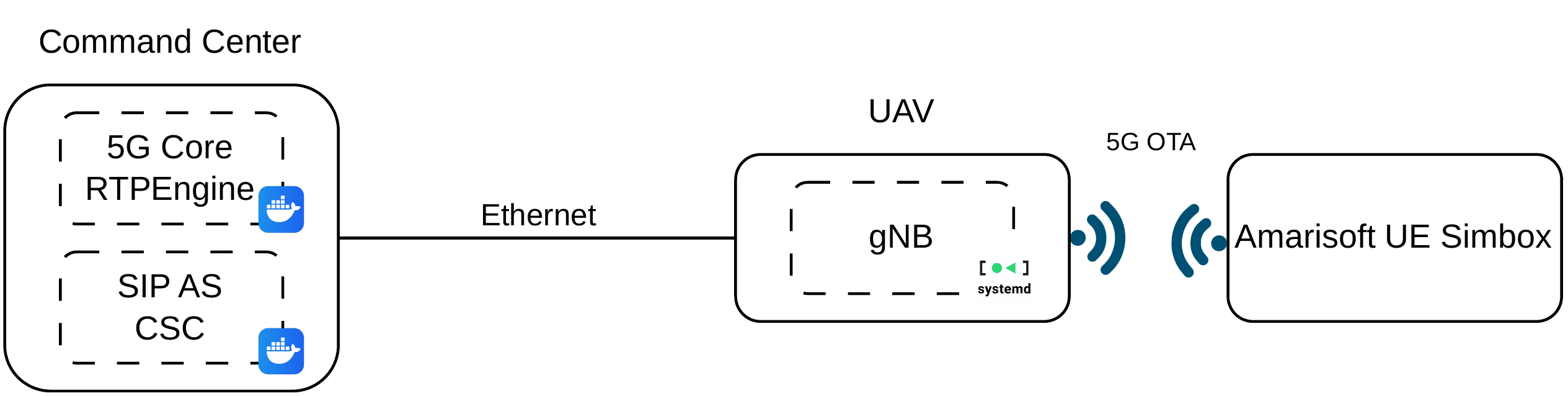}
	\caption{Deployment Configuration S1.}
	\label{fig:S1}
\end{figure}

The first configuration provides the reference condition for the subsequent experiments, depicted as S1 in Figure~\ref{fig:S1}. The Amarisoft \ac{gnb} running on the tactical node is directly connected through Ethernet to the central node. The 5G Core, \ac{upf}, \ac{ims}, Common Services Core, \ac{sip} Application Server, and RTPengine are placed on the central command center. Consequently, both signaling and user plane traffic go through the local Ethernet connection without an \ac{ntn} segment. This configuration is used for baseline functional and performance validation of \ac{mcptt}, \ac{mcvideo}, and \ac{mcdata}, as well as for the initial evaluation of signaling latency, media performance, scalability, concurrent-service operation and cross-layer priority mechanisms.

\begin{figure}[H]
	\centering
	\includegraphics[width=1\columnwidth]{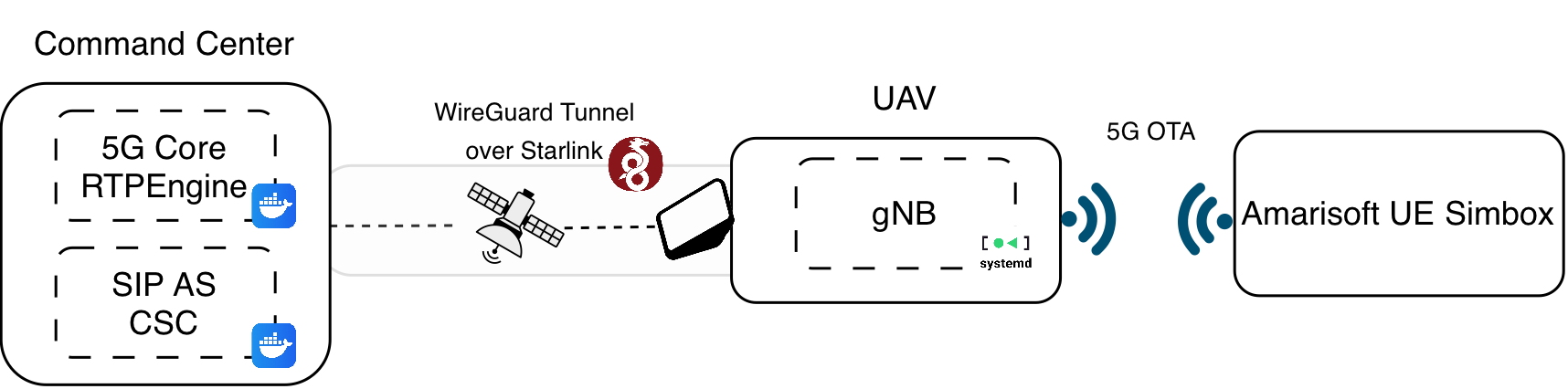}
	\caption{Deployment Configuration S2.}
	\label{fig:S2}
\end{figure}

The second configuration introduces the \ac{ntn} backhaul while preserving the centralized placement of the network and service functions, depicted as S2 in Figure~\ref{fig:S2}. The tactical \ac{gnb} is connected through Starlink to the central node, where the \ac{upf}, RTPengine, 5G Core control functions, \ac{ims}, Common Services Core, and \ac{sip} Application Server remain deployed. In this configuration, both mission-critical signaling and user plane traffic depend on the non-terrestrial backhaul. In particular, media packets must traverse the Starlink path between the \ac{gnb} and the centrally located \ac{upf}, and media relay before being forwarded toward their destination. This configuration, therefore, characterizes the performance penalty introduced when the complete mission-critical traffic path remains centrally anchored across the external backhaul.

\begin{figure}[H]
	\centering
	\includegraphics[width=1\columnwidth]{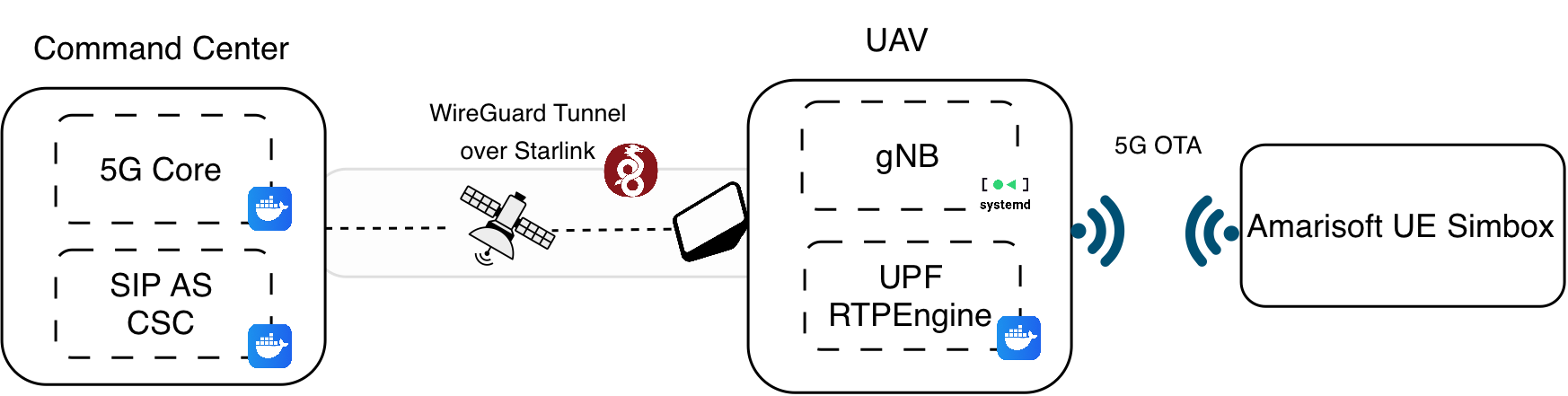}
	\caption{Deployment Configuration S3.}
	\label{fig:S3}
\end{figure}

The third configuration moves the \ac{upf} and RTPengine to the tactical node alongside the \ac{gnb}, while the 5G Core control functions, \ac{ims}, Common Services Core, and \ac{sip} Application Server remain on the central node and are reached through Starlink, depicted as S3 in Figure~\ref{fig:S3}. The tactical node, therefore, provides local user plane anchoring and media relay while retaining centralized session and service control. This placement separates the dependency of signaling from that of mission-critical media. Session establishment, service authorization, group management, and policy control signaling may traverse the \ac{ntn} backhaul, whereas media exchanged between locally connected participants can remain within the tactical network after the required state has been established. Comparing this configuration with S2 quantifies the effect of moving latency-sensitive user plane and media functions closer to the \ac{ran}, including its impact on \ac{mcptt} media delay, packet loss, jitter, \ac{mcvideo} performance, and backhaul utilization.

\begin{table}
\centering
\caption{Backhaul and function-placement scenarios used in the MCPTT evaluation.}
\label{tab:placement_scenarios}
\footnotesize
\setlength{\tabcolsep}{2.5pt}
\begin{tabular*}{\columnwidth}{@{\extracolsep{\fill}}llll@{}}
\toprule
\textbf{Scenario} & \textbf{Backhaul} & \textbf{UPF} & \textbf{RTPengine} \\
\midrule
S1 & Terrestrial & Command Center & Command Center \\
S2 & Starlink & Command Center & Command Center \\
S3 & Starlink & Tactical edge (UAV) & Tactical edge (UAV) \\
\bottomrule
\end{tabular*}
\end{table}

\begin{figure}
	\centering
	\includegraphics[width=1\columnwidth]{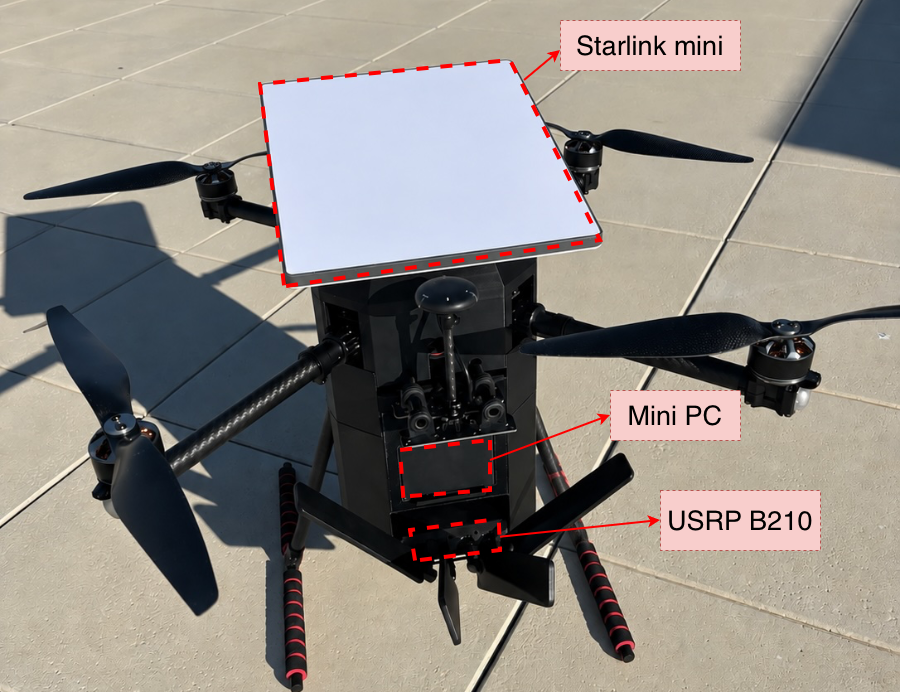}
	\caption{Drone equipped with USRP B210 and a Mini PC running Amarisoft gNB, UPF and RTPengine.}
	\label{fig:aerial_testbed}
\end{figure}

A photograph of the aerial platform and its integrated radio, computing, and non-terrestrial communication components is shown in Figure~\ref{fig:aerial_testbed}. 

\subsection{QPP Enforcement under Mission-Critical Resource Contention} 
\label{sec:priority_evaluation}

Three controlled experiments evaluated the network effect of operational priority changes under competing \ac{qos} requests: static normal priority, permanent emergency priority, and a normal $\rightarrow$ emergency $\rightarrow$ normal transition. Four simulated \acp{ue} generated uplink UDP traffic through the radio access network. Each \ac{ue} offered 15~Mbps using iperf, and requested a dedicated 5QI~65 QoS flow with uplink \ac{gbr}/\ac{mbr}, both set to 20~Mbps. The traffic rate and requested bit rates remained unchanged in all experiments. Table~\ref{tab:priority_iperf_ues} defines five operational profiles. In the static experiments, $UE_1$--$UE_3$ retain their respective profiles. The dynamic experiment instead uses $UE_1$ and two distinct senders with the $UE_3$ profile, denoted $UE_{3\mathrm{-A}}$ and $UE_{3\mathrm{-B}}$. In all experiments, sender $UE_4$ uses the $UE_0$ low-priority profile in its normal state and the $UE_4$ emergency profile when elevated.

\begin{table}
\centering
\caption{Operational priority profiles and associated network policies (Table~\ref{tab:priority_mapping}). Lower application and ARP levels indicate higher priority; \ac{pc} and \ac{pv} denote preemption capability and preemption vulnerability.}
\label{tab:priority_iperf_ues}
\footnotesize
\setlength{\tabcolsep}{2pt}
\begin{tabularx}{\columnwidth}{@{}l>{\raggedright\arraybackslash}Xcccc@{}}
\toprule
\textbf{Profile} & \textbf{Operational class} & \textbf{App. PL} & \textbf{ARP PL} & \textbf{PC} & \textbf{PV} \\
\midrule
$UE_4$ & Emergency & 1 & 1 & Yes & No \\
$UE_3$ & Incident commander & 2 & 3 & Yes & Yes \\
$UE_2$ & Team leader & 3 & 5 & No & Yes \\
$UE_1$ & Responder & 4 & 7 & No & Yes \\
$UE_0$ & Low priority & 5 & 9 & No & Yes \\
\bottomrule
\end{tabularx}
\end{table}

Application priority determines the operational class used by the service layer, including floor arbitration. At the network layer, the corresponding ARP level and \ac{pc}/\ac{pv} flags govern admission and preemption. A lower ARP level gives a request higher priority, but preempting an existing vulnerable flow also requires the new request to have \ac{pc} enabled.

To isolate network policy control from \ac{mcx} service signaling, a dedicated test harness submitted N5 requests directly to the \ac{pcf}. \ac{ngap} records were examined for dedicated-flow admission and preemption events. Uplink UDP throughput was measured on N3 and classified as 5QI~65 or 5QI~9 using the corresponding QFI mappings. The resulting evidence pertains to network policy control; it does not establish the behavior of the complete MCX emergency activation path or application-level media delivery.

\subsubsection{Experimental Sequence}

All experiments used competitor requests scheduled at 5, 45, and 90~s after sender $UE_4$ started at 0~s. The static experiments used $UE_1$, $UE_2$, and $UE_3$, respectively. The dynamic experiment used $UE_1$, $UE_{3\mathrm{-A}}$, and $UE_{3\mathrm{-B}}$ at the same times. The latter two are separate attached \acp{ue} with identical $UE_3$ policies (ARP level~3, \ac{pc} enabled, \ac{pv} enabled), offered rates, and requested bit rates. The $UE_2$ operational profile is not used in the dynamic run. This sequence tests a preemption-capable contender during emergency protection and another after normal priority is restored. Each initial policy request was submitted through an N5 \texttt{POST}, and session cleanup was scheduled at 120~s. These times are relative to the scenario start; the throughput plots use the first received traffic as their origin.

Sender $UE_4$'s assigned profile distinguished the experiments. It retained the $UE_0$ profile (ARP level~9, \ac{pc} disabled, \ac{pv} enabled) in the static-normal experiment, and the $UE_4$ profile (ARP level~1, \ac{pc} enabled, \ac{pv} disabled) in the permanent emergency experiment. In the dynamic experiment, $UE_4$ changed between these two profiles during the run.

\subsubsection{Static Normal Priority}

\begin{figure}[htbp]
\centering
\includegraphics[width=\columnwidth]{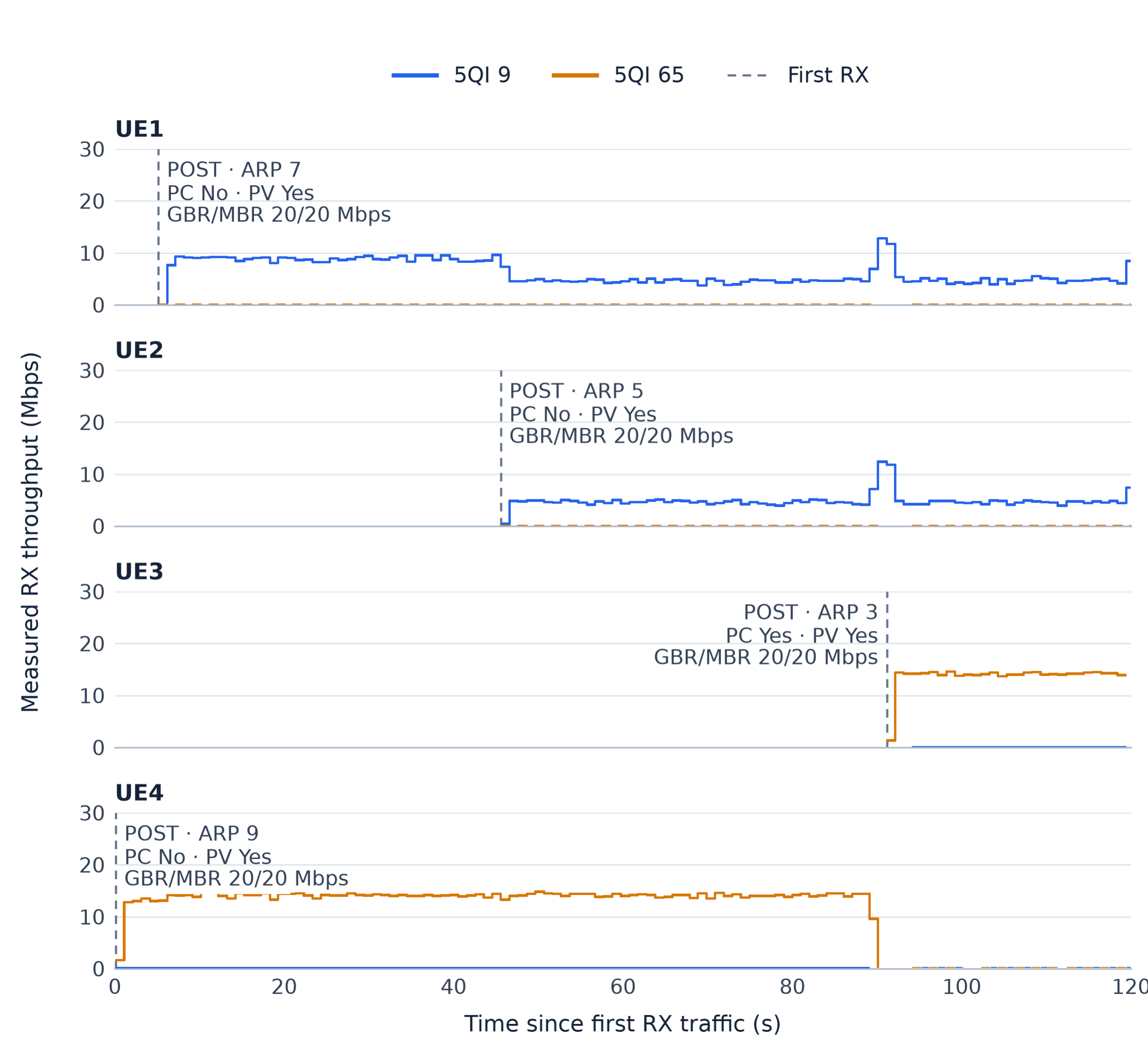}
\caption{Static normal priority. $UE_1$ and $UE_2$ transmit over 5QI~9; $UE_3$ obtains a 5QI~65 flow at approximately 90~s and preempts $UE_4$'s 5QI~65 flow. Each panel identifies a traffic sender; line colors distinguish 5QI~9 from 5QI~65. Vertical dashed markers indicate the first received traffic for each sender. Time is measured from the first received traffic in the experiment.}
\label{fig:priority_normal}
\end{figure}

$UE_4$'s initial 20~Mbps \ac{gbr} request was admitted with the $UE_0$ profile. When $UE_1$ and $UE_2$ joined at 5 and 45~s, their N5 \texttt{POSTs} were accepted by the \ac{pcf}, but the \ac{gnb} rejected their dedicated flow additions with the \ac{ngap} cause \texttt{radio-resources-not-available}. The refusals therefore occurred during radio admission, after policy acceptance. Both contenders requested the same 20~Mbps \ac{gbr} as $UE_4$ and had \ac{pc} disabled. Although their \ac{arp} levels ranked above $UE_4$'s, neither could displace its already admitted, vulnerable flow to obtain resources. $UE_1$ and $UE_2$ continued transmitting over 5QI~9, as shown in Figure~\ref{fig:priority_normal}. Their dedicated-flow refusals therefore concerned admission of the requested GBR resources, while their existing best-effort connectivity remained available.

$UE_3$ introduced a different admission outcome at 90~s. Its profile combined a higher ARP priority than $UE_4$'s normal profile with \ac{pc} enabled, allowing it to preempt $UE_4$'s vulnerable flow. NGAP records explicitly identify $UE_4$'s release as preemption, and the figure shows UE4's dedicated-flow traffic ceasing as UE3 begins transmitting over its newly admitted 5QI~65 flow. This case illustrates why a higher priority level alone was insufficient for $UE_1$ and $UE_2$: $UE_3$ additionally had permission to preempt an eligible flow.

$UE_4$'s iperf sender remained active after preemption, but no positive 5QI~9 throughput was observed for $UE_4$. Continuity over the default flow was therefore not demonstrated\footnote{In this deployment, the Open5GS AMF logged ''Not implemented'' for the NGAP \texttt{PDUSessionResourceNotify} procedure used by the gNB to report the released QFI. Subsequent signaling attempted to re-establish that QFI and encountered a PDU-session resource setup failure. These observations limit conclusions about traffic continuity after RAN-initiated preemption.}.

\subsubsection{Permanent Emergency Priority}

\begin{figure}[htbp]
\centering
\includegraphics[width=\columnwidth]{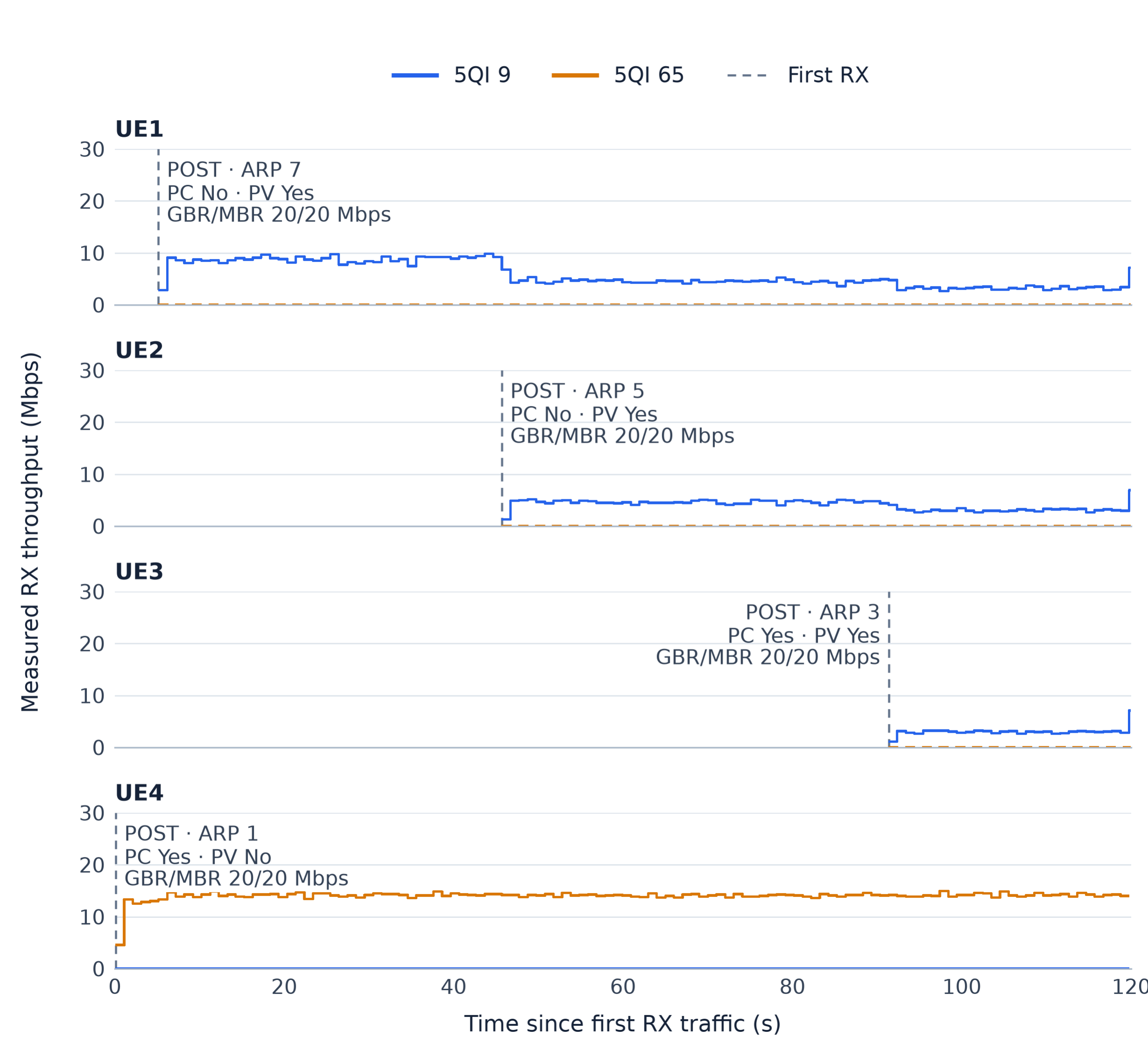}
\caption{Permanent emergency priority. $UE_4$ retains its 5QI~65 flow after all competitor arrivals; $UE_1$--$UE_3$ transmit over 5QI~9. Panels, line colors, and vertical dashed markers have the same meanings as in Figure~\ref{fig:priority_normal}.}
\label{fig:priority_emergency}
\end{figure}

With the $UE_4$ emergency profile applied from the start, $UE_4$ retained its dedicated flow throughout the experiment (Figure~\ref{fig:priority_emergency}). The \ac{pcf} accepted all four initial N5 requests, but the gNB rejected the dedicated-flow additions for $UE_1$, $UE_2$, and $UE_3$ with \texttt{radio-resources-not-available}. As in the normal case, $UE_1$ and $UE_2$ could not preempt. $UE_3$ had \ac{pc} enabled, but $UE_4$ now had both a higher ARP priority and \ac{pv} disabled. Consequently, $UE_3$ could not obtain its requested 20~Mbps GBR by replacing $UE_4$'s flow, and no preemption was recorded.

The competing \acp{ue} continued transmitting over 5QI~9. Compared with the normal case, the decisive difference occurs at $UE_3$'s arrival: the same preemption-capable contender is unable to displace $UE_4$ under the emergency profile. The observed outcome is retention of the established dedicated flow despite subsequent competing GBR requests.

\subsubsection{Dynamic Emergency Priority}

$UE_4$ started with the $UE_0$ profile; an N5 \texttt{PATCH} applied the emergency profile at 30~s, and a second \texttt{PATCH} restored normal priority at 75~s. Both updates modified the existing \ac{qos} flow while preserving \ac{5qi}~65 and the requested uplink \ac{gbr}/\ac{mbr} of 20/20~Mbps.

$UE_4$'s initial dedicated flow was admitted, while $UE_1$'s request was rejected by the \ac{gnb} for unavailable radio resources. At 45~s, $UE_{3\mathrm{-A}}$ requested a dedicated flow with the $UE_3$ profile. Although this profile had \ac{pc} enabled, $UE_4$'s emergency profile had higher priority and \ac{pv} disabled. The \ac{gnb} rejected the request with \texttt{radio-resources-not-available}, and $UE_{3\mathrm{-A}}$ continued transmitting over 5QI~9 while $UE_4$ retained its \ac{5qi}~65 flow.

\begin{figure}
\centering
\includegraphics[width=\columnwidth]{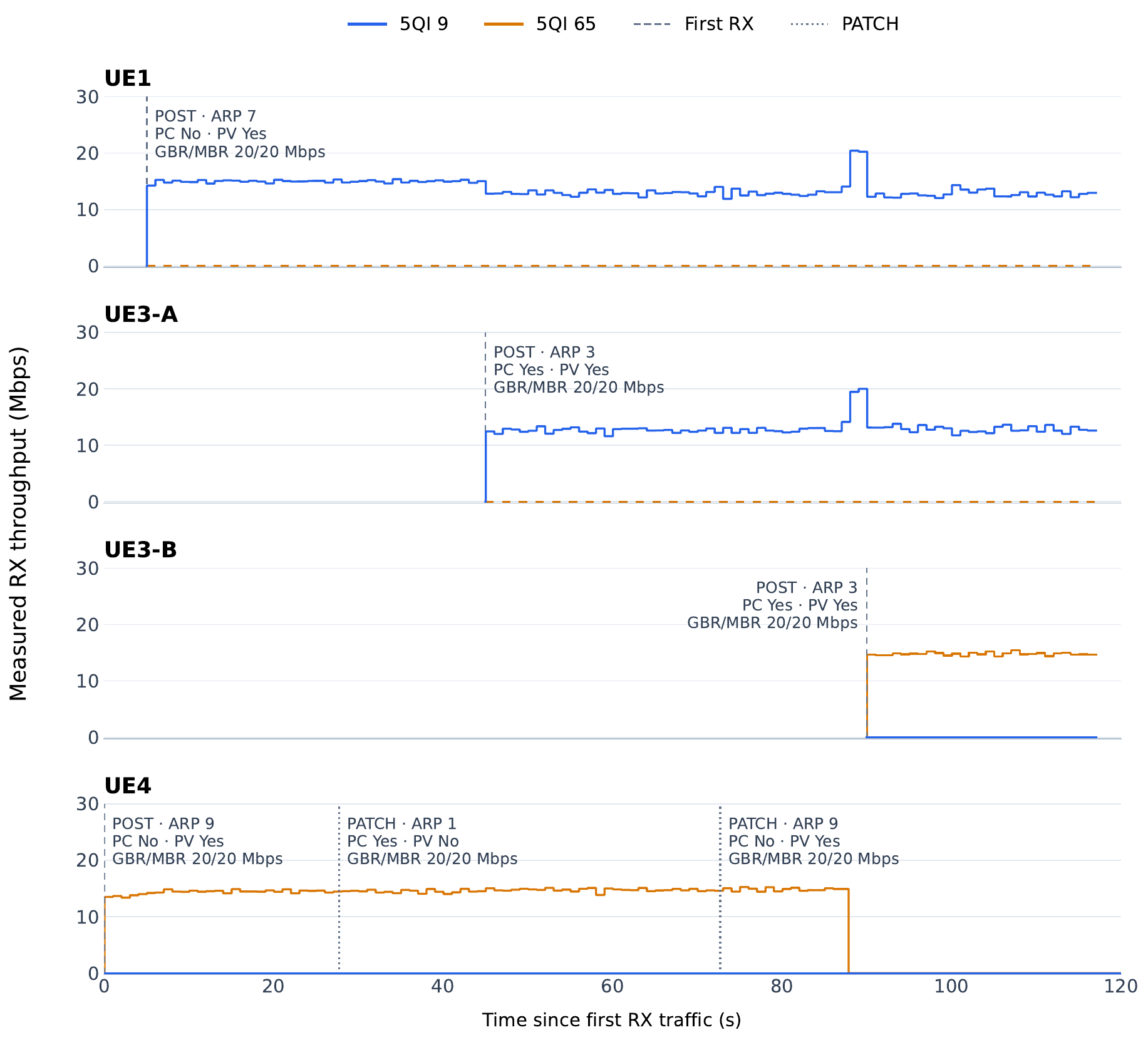}
\caption{Dynamic emergency priority. $UE_{3\mathrm{-A}}$ and $UE_{3\mathrm{-B}}$ use the same $UE_3$ profile. $UE_{3\mathrm{-A}}$ remains on 5QI~9 while $UE_4$ is protected; $UE_{3\mathrm{-B}}$ obtains 5QI~65 and preempts $UE_4$ after normal priority is restored.}
\label{fig:priority_dynamic}
\end{figure}

Restoring normal priority at 75~s made $UE_4$'s flow vulnerable to preemption but did not release it. At 90~s, $UE_{3\mathrm{-B}}$ requested a dedicated flow with the same profile as $UE_{3\mathrm{-A}}$. Its 5QI~65 flow was admitted and preempted $UE_4$'s flow, as confirmed by \ac{ngap} records and the traffic traces in Fig.~\ref{fig:priority_dynamic}. $UE_{3\mathrm{-A}}$ remained on 5QI~9. The two requests therefore produced different admission outcomes before and after emergency protection was removed.

Together, these runs show how the priority profiles affect admission and retention under competing \ac{gbr} requests. $UE_3$ preempted $UE_4$'s normal flow in the static experiment, but could not displace its permanent emergency flow. The dynamic experiment showed the same distinction within one run: $UE_{3\mathrm{-A}}$ was refused during emergency priority, whereas $UE_{3\mathrm{-B}}$ was admitted after normal priority was restored. Each configuration was executed once, and \ac{arp} level, \ac{pc}, and \ac{pv} changed together; the experiments do not isolate each attribute's contribution or establish admission probabilities.

\subsubsection{UE-to-RAN QoS signaling Latency}

This experiment measures the interval from \ac{ue} \ac{sip} socket submission, through \ac{ims}, \ac{sip} \ac{as}, and N5/Core signaling, to the matching successful \ac{ngap} \ac{pdu} \texttt{Session Resource Modify Response} captured at the Core host. S1 and S3 each used one \ac{ue} without competing flows or media traffic. In each scenario, 500 affiliation requests generated 500 N5 \texttt{POSTs}. A separate series of 250 floor-request/floor-release pairs generated 250 emergency-request \texttt{PATCHes} and 250 emergency-release \texttt{PATCHes} on an established \ac{qfi}, yielding 1,000 measured operations per scenario. The emergency transitions were exercised through \ac{sip} floor operations with \ac{cms} priority prepared before timing; these \texttt{PATCHes} also changed direction and bit rates. All 1,000 operations per scenario matched successful \ac{ngap} responses, and \texttt{PATCHes} preserved \ac{qfi} and \ac{5qi}.

\begin{figure}[htbp]
\centering
\includegraphics[width=\columnwidth]{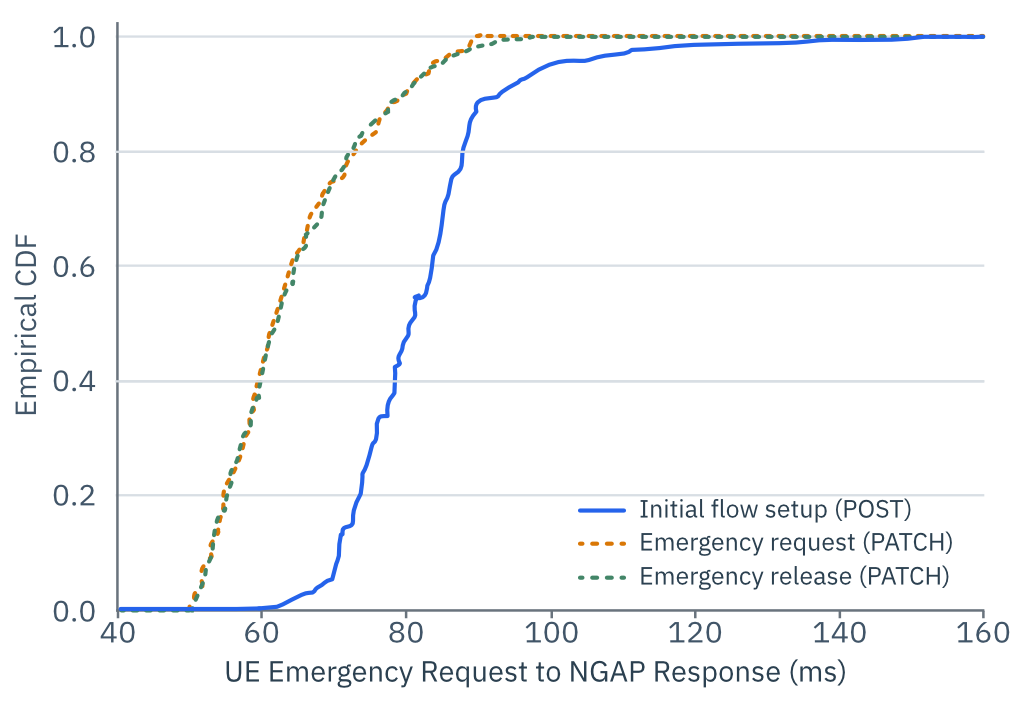}
\caption{S1: UE SIP submission to the matching successful NGAP response. Initial flow setup is affiliation (N5 POST, $n=500$); the plotted emergency request and release are floor request and release with prepared CMS priority (N5 PATCH, $n=250$ each).}
\label{fig:priority_ue_ngap_s1}
\end{figure}

\begin{figure}[htbp]
\centering
\includegraphics[width=\columnwidth]{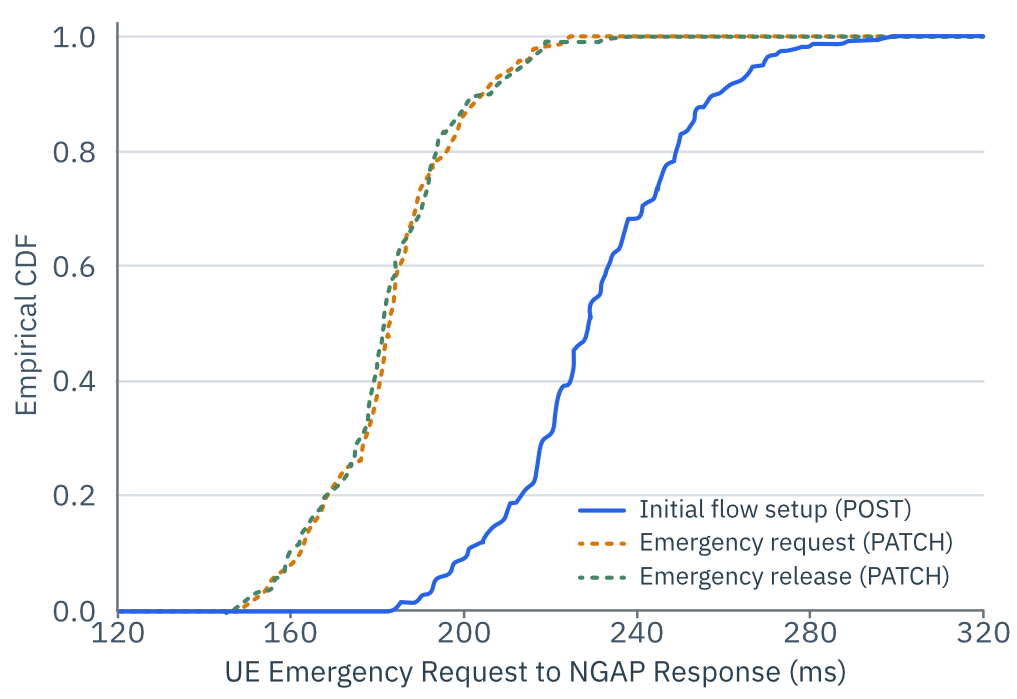}
\caption{S3: the same UE-to-NGAP signaling interval and operations as in Figure~\ref{fig:priority_ue_ngap_s1}. The displayed POST CDF omits four observations above 350~ms and is normalized over the remaining 496; the full-sample POST maximum is 1248.41~ms. All 500 POST observations are retained in the reported percentiles.}
\label{fig:priority_ue_ngap_s3}
\end{figure}

For initial flow setup, emergency request, and emergency release, respectively, S1 median latencies were 80.5, 61.7, and 61.8~ms, with $P_{95}$ values of 98.8, 83.3, and 83.6~ms (Figure~\ref{fig:priority_ue_ngap_s1}). In S3, the medians increased to 229.0, 182.6, and 181.7~ms, and $P_{95}$ to 267.8, 211.7, and 212.6~ms (Figure~\ref{fig:priority_ue_ngap_s3}). Initial flow setup took longer than updates to an existing flow, while emergency-request and emergency-release distributions nearly overlapped. Per-operation clock-offset calibration bounded the maximum retained interval width to 0.72~ms in S1 and 1.57~ms in S3.

The higher S3 latencies are consistent with service and Core--\ac{gnb} signaling continuing across Starlink despite edge-local \ac{upf} and media functions; the comparison does not isolate backhaul delay from processing. All observations contribute to the reported statistics. These measurements quantify signaling completion, excluding registration, initial \ac{pdu}-session establishment, and media delivery; they do not identify the precise instant of scheduler enforcement.

\subsection{MCPTT Performance and Latency Analysis}

This subsection characterizes the performance of the implemented \ac{mcptt} service, with particular emphasis on floor control latency, access time, and media transport. The measurements distinguish application-level floor decisions from the subsequent media path response, allowing the contribution of signaling and media transport to be evaluated separately across the considered deployment configurations.

\subsubsection{Floor Control Performance}

In half-duplex \ac{mcptt}, floor control determines which group member is authorized to transmit at a given time. The experiment evaluates floor acquisition, refusal, release, reacquisition and priority preemption. In S3, five affiliated \acp{ue} participated: four used the $UE_3$ (Incident Commander) profile, with application priority level~2, and one used the $UE_2$ (Team Leader) profile, with application priority level~3. Lower values indicate higher priority.

Floor control response latency is measured from the moment a \ac{ue} sends a floor request until it receives the final \ac{sip} response. One participant with the $UE_3$ profile is used to measure initial acquisition and reacquisition. Equal-priority refusal is measured when another $UE_3$ profile participant requests the floor while the first holds it. For preemption, a higher-priority $UE_3$ profile participant requests the floor while the $UE_2$ profile participant holds it. A subsequent request from the $UE_2$ profile participant is refused. Release latency is measured for both profiles.

Two additional measurements begin with the $UE_3$ profile participant's preemption request. Revoke notification latency ends when the $UE_2$ profile participant receives the \texttt{floor revoked} notification. First new-talker \ac{rtp} latency ends when a listening $UE_3$ profile participant receives the first \ac{rtp} packet from the new talker. These measurements distinguish floor response, notification delivery, and media arrival. Table~\ref{tab:floor_control_latency} reports the $P_{50}$ and $P_{95}$ values for the three placement scenarios.

\begin{table}
\centering
\caption{Floor control latency by placement scenario. Values are in milliseconds.}
\label{tab:floor_control_latency}
\scriptsize
\setlength{\tabcolsep}{2pt}
\begin{tabularx}{\columnwidth}{@{}>{\raggedright\arraybackslash}Xrr@{\hspace{8pt}}rr@{\hspace{8pt}}rr@{}}
\toprule
& \multicolumn{2}{c}{\textbf{S1}} & \multicolumn{2}{c}{\textbf{S2}} & \multicolumn{2}{c}{\textbf{S3}} \\
\cmidrule(lr){2-3}\cmidrule(lr){4-5}\cmidrule(l){6-7}
\textbf{Measurement} & \textbf{$P_{50}$} & \textbf{$P_{95}$} & \textbf{$P_{50}$} & \textbf{$P_{95}$} & \textbf{$P_{50}$} & \textbf{$P_{95}$} \\
\midrule
Initial acquisition       & 53.36 & 53.70 & 93.37  & 95.24  & 86.89  & 91.83  \\
Equal-priority refusal    & 44.58 & 48.54 & 95.58  & 102.79 & 89.64  & 99.57  \\
Reacquisition             & 35.67 & 43.24 & 91.35  & 94.93  & 88.50  & 92.43  \\
Priority preemption       & 46.14 & 46.75 & 99.55  & 104.54 & 93.90  & 108.07 \\
Revoke notification       & 44.64 & 44.75 & 95.65  & 104.08 & 90.12  & 100.53 \\
First new-talker RTP      & 79.65 & 85.08 & 163.08 & 194.39 & 106.95 & 126.32 \\
Lower-priority refusal    & 44.58 & 44.68 & 98.10  & 100.58 & 81.65  & 88.24  \\
Floor release             & 35.22 & 38.41 & 77.67  & 101.44 & 84.55  & 92.39  \\
\bottomrule
\end{tabularx}
\end{table}

\begin{figure}
    \centering
    \includegraphics[width=\columnwidth]{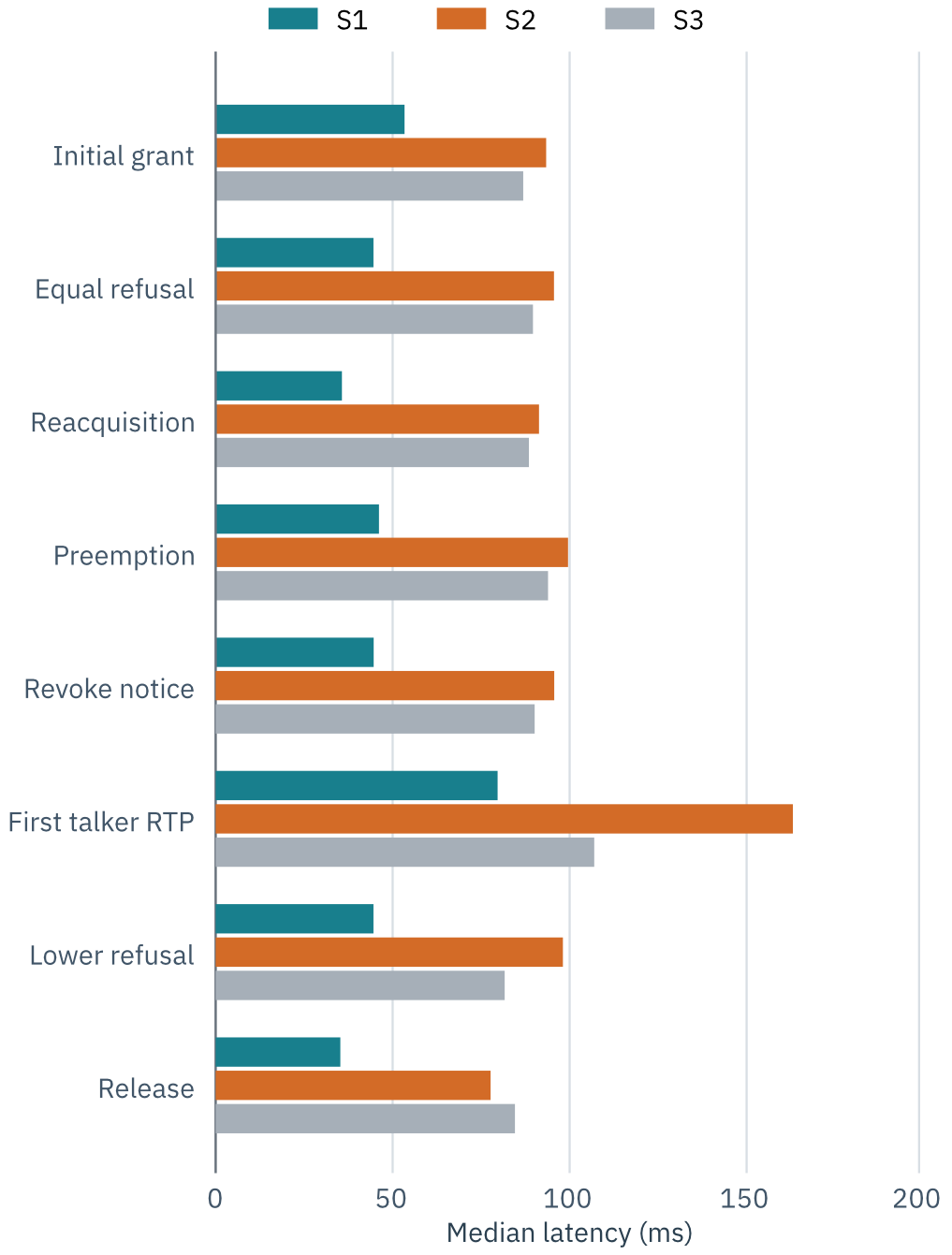}
    \caption{Median UE-observed floor control latency by operation and placement scenario.}
    \label{fig:mcptt_floor_control}
\end{figure}

The S3 run completed all five rounds, and all 1,000 floor requests received their expected responses. No timeout or missing reply was observed. Figure~\ref{fig:mcptt_floor_control} compares the median latencies. Relative to the Ethernet baseline S1, S2 increased every \ac{ue} observed metric. In S2, the \ac{sip} \acl{as}, \ac{upf}, and RTPengine remained at the command center, so floor control signaling and media depended on the Starlink backhaul. The largest increase was for first new-talker \ac{rtp}: 83.43~ms at $P_{50}$ and 109.31~ms at $P_{95}$.

In S3, the \ac{sip} response medians remained broadly similar to those in S2, consistent with the \ac{sip} \acl{as} remaining at the command center and being reached over Starlink in both scenarios. Five of the six medians were lower in S3, while the release median was 6.88~ms higher; the revoke notification median was 5.53~ms lower. First new-talker \ac{rtp} fell from 163.08 to 106.95~ms at $P_{50}$ and from 194.39 to 126.32~ms at $P_{95}$, reductions of 56.13 and 68.07~ms, respectively. This improvement is consistent with placing the \ac{upf} and RTPengine on the \ac{uav}, allowing media to be relayed locally instead of travelling to the command center and back. The interval still includes floor control signaling to the centralized \ac{sip} \acl{as}, which helps explain why S3 remains slower than S1.

\subsubsection{MCPTT Access Time}

\ac{3gpp} defines \ac{mcptt} Access Time (KPI~1) as the interval between a user's request to speak and the indication that transmission may begin. For an already established call, \ac{3gpp} TS~22.179 requires this interval to remain below 300~ms for 95\% of \ac{mcptt} \ac{ptt} requests~\cite{3gpp_ts_22_179}. In our evaluation, this protocol level interval was measured from transmission of the \texttt{floor-request} to reception of the final \ac{sip} response by 64 pre-affiliated simulated \acp{ue} following a relay pattern. Table~\ref{tab:mcptt_access_time} reports the resulting $P_{50}$, $P_{95}$, $P_{99}$, and maximum access times for the three deployment scenarios.

\begin{table}
\centering
\caption{Protocol-level \ac{mcptt} access time by placement scenario.}
\label{tab:mcptt_access_time}
\small
\setlength{\tabcolsep}{4pt}
\begin{tabular*}{\columnwidth}{@{\extracolsep{\fill}}lrrrr@{}}
\toprule
\textbf{Scenario} & \textbf{$P_{50}$} & \textbf{$P_{95}$} & \textbf{$P_{99}$} & \textbf{Max} \\
\midrule
S1 & 47.84 & 67.16 & 72.53 & 148.73 \\
S2 & 90.78 & 114.24 & 128.71 & 355.98 \\
S3 & 149.25 & 184.52 & 212.71 & 341.63 \\
\bottomrule
\end{tabular*}
\end{table}

No timeout or worker failure was observed in any of the three scenarios. Relative to the terrestrial baseline S1, introducing the Starlink backhaul with centralized service, user plane, and media functions in S2 increased the median access time from 47.84~ms to 90.78~ms and the $P_{95}$ from 67.16~ms to 114.24~ms, corresponding to increases of 42.94~ms and 47.08~ms respectively. In S3, where the \ac{upf} and RTPengine were moved to the tactical edge while the \ac{sip} \ac{as} and other service control functions remained centralized, the access time $P_{50}$ and $P_{95}$ further increased to 149.25~ms and 184.52~ms, respectively. 

\begin{figure}
    \centering
    \includegraphics[width=\columnwidth]{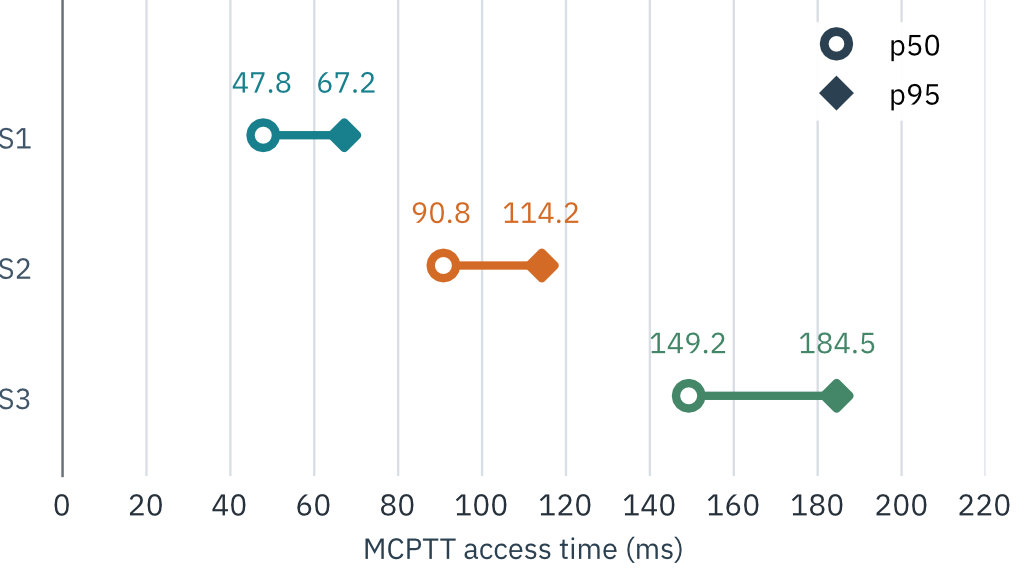}
    \caption{Protocol-level MCPTT access time by placement scenario, showing the median and 95th percentile.}
    \label{fig:mcptt_access_time}
\end{figure}

Despite these differences, the measured $P_{95}$ remained below the 300~ms KPI1 reference in all three scenarios, as illustrated in Figure~\ref{fig:mcptt_access_time}. Isolated observations above 300~ms occurred in S2 and S3, producing maxima of 355.98~ms and 341.63~ms respectively. 

\subsubsection{Audio Transport Performance}

\ac{mcptt} audio transport was evaluated using 64 affiliated \acp{ue}, with one talker transmitting repeated five second \ac{rtp} audio bursts to 63 listeners at 20~ms packet intervals. Transport delay was measured from submission of each packet at the sender socket to its reception callback at the listener, thereby capturing the communication path through the 5G network and media relay. Table~\ref{tab:mcptt_audio_transport} reports the resulting delay percentiles and packet loss pooled across listeners and transmitted bursts.

\begin{table}
\centering
\caption{\ac{mcptt} audio transport results pooled across listeners and bursts.}
\label{tab:mcptt_audio_transport}
\small
\setlength{\tabcolsep}{4pt}
\begin{tabular*}{\columnwidth}{@{\extracolsep{\fill}}lrrrr@{}}
\toprule
\textbf{Scenario} & \textbf{$P_{50}$} & \textbf{$P_{95}$} & \textbf{$P_{99}$} & \textbf{Loss} \\
\midrule
S1 & 39.22 & 60.64 & 69.44 & 0.73\% \\
S2 & 86.00 & 117.53 & 142.44 & 9.78\% \\
S3 & 39.19 & 57.50 & 62.62 & 0.36\% \\
\bottomrule
\end{tabular*}
\end{table}

All workers completed successfully, and every burst/listener measurement was valid in S1, S2, and S3. The delay percentiles are calculated from the raw received packet observations, whereas packet loss is calculated relative to the total number of packets expected across all listeners. Centralized media anchoring over the Starlink backhaul in S2 substantially degraded both metrics relative to the terrestrial baseline S1: the median transport delay increased from 39.22~ms to 86~ms, the $P_{95}$ increased from 60.64~ms to 117.53~ms, and packet loss increased from 0.73\% to 9.78\%.

Moving the media relay to the tactical edge in S3 substantially changed this behavior. The median delay decreased from 86~ms in S2 to 39.19~ms in S3, while the $P_{95}$ decreased from 117.53~ms to 57.50~ms and packet loss from 9.78\% to 0.36\%. S3 therefore recovered transport performance comparable to, and in these runs slightly better than, the terrestrial S1 baseline despite retaining the non-terrestrial backhaul for centralized service functions. This contrast indicates that the degradation observed in S2 is primarily associated with carrying mission-critical media through the centralized Starlink path, whereas edge local media anchoring avoids this traversal for \ac{mcptt} traffic.

\begin{figure}
    \centering
    \includegraphics[width=\columnwidth]{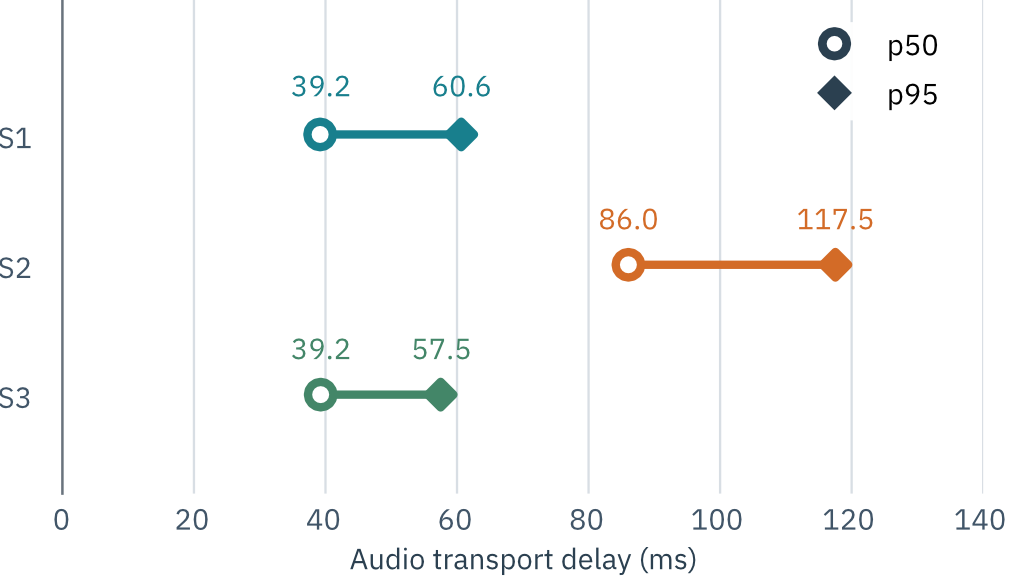}
    \caption{\ac{mcptt} audio transport delay by placement scenario, showing the median and 95th percentile of received packets.}
    \label{fig:mcptt_audio_transport}
\end{figure}

Figure~\ref{fig:mcptt_audio_transport} highlights this effect by comparing the median and 95th-percentile transport delays across the three placement scenarios.

\subsubsection{Central-host Processing Intervals}

To assess whether the \ac{ue}-observed delay arose within centralized signaling, we captured both \ac{sip} legs of the \ac{pcscf} on the central-host with host timestamps and instrumented the \ac{sip} \ac{as} with monotonic timestamps. The SimBox separately measured the end-to-end floor control transaction time, from submission of a floor control request to reception of the corresponding final response, using a \ac{ue}-side monotonic clock. The central-host analysis reassembled TCP streams and matched each request to its response using the SIP Call-ID, CSeq, and dialog tags. Table~\ref{tab:mcptt_components} reports $P_{95}$ intervals for one representative operation per experiment: free-floor acquisition for access time and audio transport, and priority preemption for floor control.

\ac{pcscf} request forwarding runs from capture of the complete \ac{ue} request to the first packet forwarded toward the \ac{as}; response forwarding runs from capture of the complete \ac{as} response to the first packet sent toward the \ac{ue}. The complete \ac{pcscf} request--response interval runs from the complete \ac{ue} request to the complete outgoing response. \ac{sip} \ac{as} processing runs from the instrumented completion of request reading to completion of final response writing, including awaited work such as the \ac{cms} priority lookup during preemption. The \ac{as} and packet-capture intervals use separate clocks, and each $P_{95}$ is calculated independently; these percentiles should not be added.

\begin{table}
\centering
\caption{Central-host-observed $P_{95}$ \ac{sip} signaling intervals by \ac{mcptt} experiment. Values in milliseconds.}
\label{tab:mcptt_components}
\scriptsize
\setlength{\tabcolsep}{4pt}
\begin{tabularx}{\columnwidth}{@{}>{\raggedright\arraybackslash}Xrrr@{}}
\toprule
\textbf{Observed interval} & \textbf{S1} & \textbf{S2} & \textbf{S3} \\
\midrule
\multicolumn{4}{@{}l}{\textit{Access time}} \\
P-CSCF request forwarding  & 0.57 & 0.56 & 0.59 \\
SIP AS processing           & 4.85 & 6.07 & 8.05 \\
P-CSCF response forwarding & 0.30 & 0.25 & 0.25 \\
SIP request--response at P-CSCF & 5.66 & 6.93 & 8.96 \\
\addlinespace
\multicolumn{4}{@{}l}{\textit{Floor control}} \\
P-CSCF request forwarding  & 0.43 & 0.41 & 0.50 \\
SIP AS processing           & 6.71 & 7.41 & 7.35 \\
P-CSCF response forwarding & 0.19 & 0.16 & 0.25 \\
SIP request--response at P-CSCF & 7.33 & 8.12 & 8.10 \\
\addlinespace
\multicolumn{4}{@{}l}{\textit{Audio transport}} \\
P-CSCF request forwarding  & 0.49 & 0.47 & 0.49 \\
SIP AS processing           & 29.58 & 38.22 & 49.35 \\
P-CSCF response forwarding & 0.28 & 0.27 & 0.31 \\
SIP request--response at P-CSCF & 30.34 & 38.93 & 50.04 \\
\bottomrule
\end{tabularx}
\end{table}

Across S1--S3, each \ac{pcscf} forwarding interval remained below 0.6~ms at $P_{95}$. For access time, the \ac{ue}-observed $P_{95}$ increased from 67.16 to 114.24 and 184.52~ms, whereas the \ac{pcscf} request--response $P_{95}$ increased only from 5.66 to 6.93 and 8.96~ms. For priority preemption, the \ac{ue}-observed $P_{95}$ rose from 46.75~ms in S1 to 104.54~ms in S2, while the \ac{pcscf} interval changed from 7.33 to 8.12~ms. Thus, processing in the centralized node \ac{sip} path was not the dominant bottleneck for these \ac{ue}-observed signaling delays or their increase across the compared scenarios. The independently calculated percentiles cannot be subtracted to assign the remaining delay to a specific network segment.

The floor control \ac{pcscf} totals remained similar across all three scenarios (7.33--8.12~ms). In the audio-transport experiment, the floor-grant total increased from 30.34 to 38.93 and 50.04~ms, principally alongside SIP AS processing; these rows measure the grant before each burst, not \ac{rtp} delivery. In the S1--S2 runs, background N5 direction reconciliation had $P_{95}$ values of 11.35--12.51~ms. It was not awaited by the SIP response and confirms only \ac{pcf} acceptance, not \ac{smf}, \ac{upf}, or \ac{ran} enforcement.

\subsection{MCPTT Audio Quality}

\ac{mcptt} audio quality was evaluated using 63 receiving \acp{ue} over ten 5s Opus transmission bursts per scenario, with a 20~ms packetization interval and 157,500 expected packet deliveries. Packet-level measurements were obtained by correlating matched sender and receiver logs, allowing packet loss, one-way transport delay, interarrival jitter, and received payload bitrate to be characterized at the receiving \acp{ue}. The measured packet delivery conditions were subsequently used to derive receiver-side parametric \ac{mos} estimates under different playout buffer assumptions. The analysis therefore first characterizes the observed \ac{rtp} transport performance and then evaluates its expected impact on voice quality.

\subsubsection{Packet-Level Voice Performance}

Table~\ref{tab:mcptt_packet_quality} summarizes the packet-level measurements across the 63 receiving \acp{ue}. Packet loss represents the fraction of expected \ac{rtp} packets that were not received, while $D_{50}$ and $D_{95}$ characterize the one-way transport delay. The $J_{95}$ metric represents the 95th percentile of the RFC~3550 interarrival jitter estimate. The received Opus payload bitrate was calculated over 50~s of active transmission per listener.

\begin{table}[!t]
\centering
\caption{\ac{mcptt} packet delivery at the receiving UEs.}
\label{tab:mcptt_packet_quality}
\footnotesize
\setlength{\tabcolsep}{2pt}
\renewcommand{\arraystretch}{1.12}
\begin{tabular*}{\columnwidth}{@{}l@{\extracolsep{\fill}}rrrrrr@{}}
\toprule
\textbf{Setup} & \textbf{Loss} & $\boldsymbol{D_{50}}$ &
$\boldsymbol{D_{95}}$ & $\boldsymbol{J_{95}}$ &
\textbf{Disc.} & \textbf{Rate} \\
\midrule 
\textbf{S1} & 0.73 & 39.21 & 60.63  & 19.50 & 1097 & 59.56 \\
\textbf{S2} & 9.78 & 85.99 & 117.51 & 17.85 & 7648 & 54.14 \\
\textbf{S3} & 0.36 & 39.19 & 57.50  & 20.16 & 261  & 59.79 \\
\bottomrule
\end{tabular*}
\par\smallskip
\parbox{\columnwidth}{\scriptsize\raggedright
Loss: missing UE deliveries (\%);
$D_{50}$/$D_{95}$: one-way delay percentiles (ms);
$J_{95}$: RFC~3550 jitter percentile (ms);
Disc.: nonconsecutive RTP sequence transitions;
Rate: received Opus payload bitrate (kb/s), excluding headers.}
\end{table}

The centralized \ac{ntn} configuration S2 exhibited the greatest packet-level degradation, with loss increasing to 9.78\% and $D_{95}$ reaching 117.51~ms, compared with 0.73\% and 60.63ms in the terrestrial S1 baseline. \ac{rtp} sequence discontinuities similarly increased from 1,097 to 7,648, while the received payload bitrate decreased from 59.56 to 54.14~kb/s.

In contrast, S3 recovered packet-level performance to approximately the terrestrial baseline, with 0.36\% loss, a $D_{50}$ of 39.19~ms, a $D_{95}$ of 57.50~ms, and a received payload bitrate of 59.79~kb/s. Interarrival jitter remained comparable across the three scenarios, with $J_{95}$ ranging from 17.85 to 20.16~ms. These results are consistent with the placement of the \ac{upf} and RTPengine at the tactical edge in S3, which prevents local \ac{rtp} traffic from traversing the satellite path used in S2. 

\subsubsection{Receiver-Side Parametric Quality Estimate}

To complement the packet-level analysis in Table~\ref{tab:mcptt_packet_quality}, receiver-side voice quality was estimated offline using VoIPmonitor. The analysis considers three jitter-buffer configurations: a fixed 50-ms buffer (F1), a fixed 200-ms buffer (F2), and an adaptive buffer with a maximum size of 500~ms. These configurations provide different tolerance to packet delay variation and therefore allow the impact of the measured transport conditions on the estimated voice quality to be examined.

Voice quality was assessed considering the main network-level factors affecting media delivery, including packet loss, end-to-end delay, jitter, and the resulting packet delivery behavior. The \ac{mos} is a long established measure of perceived voice quality, originally defined through subjective listening assessments. In this work, VoIPmonitor is used to obtain objective parametric \ac{mos} estimates from the observed \ac{rtp} packet stream rather than from the decoded audio signal. Its calculation uses packet-loss probability and loss burstiness through a formulation derived from the ITU-T G.107 E-model\footnote{\url{https://www.itu.int/rec/T-REC-G.107}}, which models the impact of transmission impairments on conversational quality.

Since packet delay variation can cause packets to arrive too late for playout and therefore become effectively lost, VoIPmonitor evaluates the \ac{rtp} streams using three simulated jitter-buffer configurations. \ac{mos} F1 uses a fixed 50ms jitter buffer and is therefore more sensitive to delay variation; \ac{mos} F2 uses a fixed 200ms buffer, tolerating greater delay variation before packets are considered too late for playout; and \ac{mos} Adapt uses an adaptive jitter buffer that can grow up to 500~ms. For the Opus traffic evaluated in this work, VoIPmonitor applies its G.711/PLC-oriented parametric model rather than an Opus-specific perceptual model. The resulting scores are therefore interpreted as estimates of transport related voice impairment and are considered together with the directly measured packet loss, delay, and jitter, rather than as perceptual measurements of the decoded Opus audio.

\begin{table}
\centering
\caption{VoIPmonitor receiver-side parametric MOS estimates.}
\label{tab:mcptt_ue_mos}
\small
\setlength{\tabcolsep}{3pt}
\renewcommand{\arraystretch}{1.12}
\begin{tabular*}{\columnwidth}{@{}l@{\extracolsep{\fill}}ccc@{}}
\toprule
\textbf{Metric} & \textbf{S1} & \textbf{S2} & \textbf{S3} \\
\midrule
\multicolumn{4}{@{}l}{\textit{Fixed 50-ms buffer (F1)}} \\
\textbf{Mean UE MOS estimate} & \textbf{4.40} & \textbf{2.28} & \textbf{4.49} \\
5th-percentile UE MOS & 4.21 & 1.80 & 4.41 \\
\midrule
\multicolumn{4}{@{}l}{\textit{Fixed 200-ms buffer (F2)}} \\
\textbf{Mean UE MOS estimate} & \textbf{4.42} & \textbf{2.45} & \textbf{4.49} \\
5th-percentile UE MOS & 4.30 & 2.10 & 4.50 \\
\midrule
\multicolumn{4}{@{}l}{\textit{Adaptive buffer (up to 500 ms)}} \\
\textbf{Mean UE MOS estimate} & \textbf{4.34} & \textbf{2.41} & \textbf{4.49} \\
5th-percentile UE MOS & 4.20 & 2.00 & 4.40 \\
\bottomrule
\end{tabular*}
\end{table}

As shown in Table~\ref{tab:mcptt_ue_mos}, S3 produced the highest mean \ac{mos} estimate for all three buffer configurations, with a value of 4.49, while S1 remained between 4.34 and 4.42 and S2 between 2.28 and 2.45. The substantially lower estimates in S2 are consistent with the higher packet loss observed in Table~\ref{tab:mcptt_packet_quality}, where 15,410 packet deliveries were missing compared with 1,153 in S1 and 565 in S3. Conversely, the similar estimates obtained for S1 and S3 are consistent with their comparatively low packet-loss rates.

These results further indicate that relocating the media path to the tactical edge in S3 can preserve receiver-side transport conditions despite the use of the \ac{ntn} backhaul for centralized functions.

\subsection{MCVideo Service}
The \ac{mcvideo} stream was encoded using H.264 at a target bitrate of 750~kbit/s. This operating point was chosen to strike an optimal balance between video quality and radio resource consumption within the bandwidth-constrained mobile testbed. Prior experimental evaluations of H.264 video have analyzed 480p encoding across a broad range of bitrates including levels around 700--800~kbit/s highlighting the inherent trade-offs between encoding bitrate, perceived quality, and bandwidth consumption~\cite{video_quality}. Consequently, 750~kbit/s was adopted as a bandwidth-efficient baseline for these experiments, facilitating scalability testing with up to 64~\acp{ue}.
\begin{figure}[htbp]
    \centering
    \includegraphics[width=\linewidth]{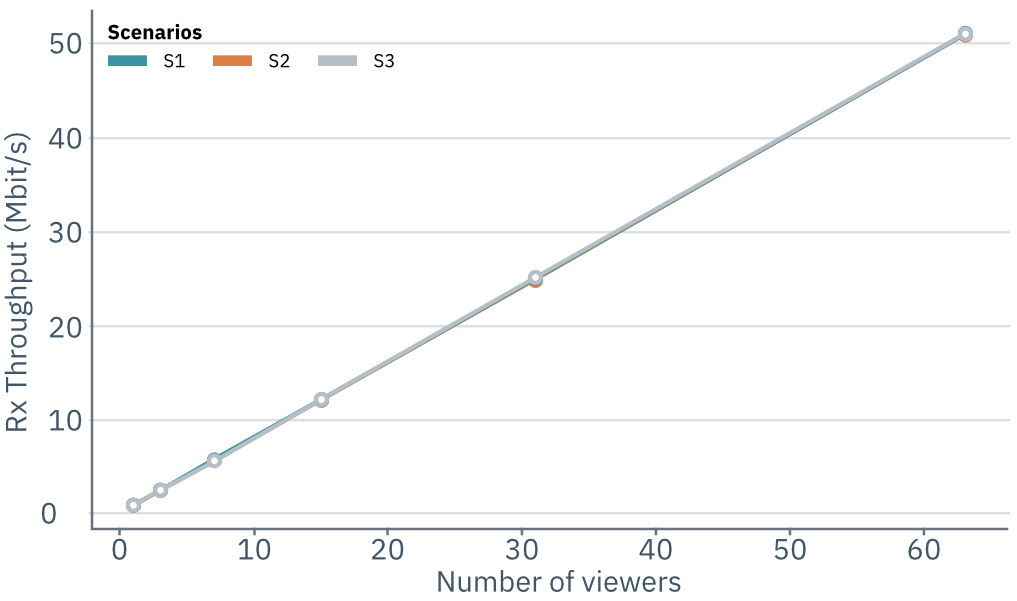}
    \caption{Total RX throughput as a function of the number of UEs.}
    \label{fig:mcvideo-rx-throughput}
\end{figure}
\begin{figure}[htbp]
    \centering
    \includegraphics[width=\linewidth]{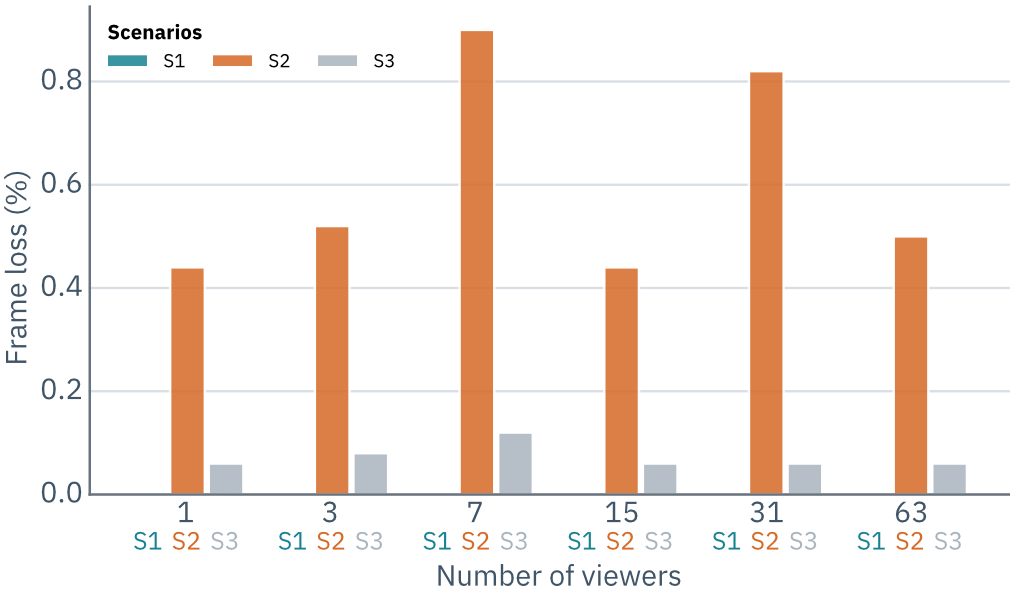}
    \caption{Frame loss as a function of the number of UEs.}
    \label{fig:mcvideo-frame-loss}
\end{figure}
\begin{figure}[htbp]
    \centering
    \includegraphics[width=\linewidth]{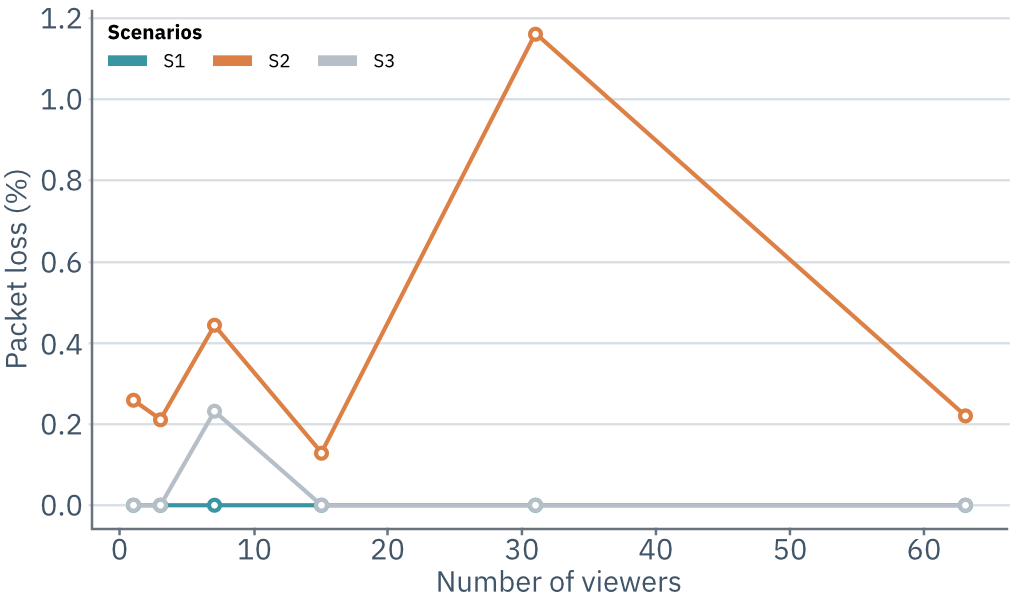}
    \caption{RTP packet loss as a function of the number of UEs.}
    \label{fig:mcvideo-rtp-loss}
\end{figure}
As depicted in Figure~\ref{fig:mcvideo-rx-throughput}, the aggregate receive throughput scales proportionally with the number of viewers. Meanwhile, Figures~\ref{fig:mcvideo-frame-loss} and \ref{fig:mcvideo-rtp-loss} reveal that, although overall performance remained stable, small but non-zero frame and RTP packet losses occurred in the NTN-backed configurations across various load levels. The system's end-to-end performance under the maximum evaluated load of 64~\acp{ue} is characterized in detail in Table~\ref{tab:mcvideo-performance-load}.

\begin{table}
    \centering
    \caption{MCVideo end-to-end performance under load (64 UEs, 1 publisher).}
    \label{tab:mcvideo-performance-load}
    \small
    \resizebox{\columnwidth}{!}{%
    \begin{tabular}{l|r r r}
        \hline
        \textbf{Evaluated Metric} & \textbf{S1} & \textbf{S2} & \textbf{S3} \\
        \hline
        \multicolumn{4}{l}{\textit{1. Signaling Latency (ms)}} \\
        Stream-establishment ($P_{50}$) & 45.28 & 88.78 & 214.86 \\
        Stream-establishment (mean) & 45.28 & 70.09 & 90.97 \\
        Viewer-attachment ($P_{50}$) & 125.58 & 204.94 & 221.11 \\
        Viewer-attachment (mean) & 140.25 & 203.20 & 214.04 \\
        \hline
        \multicolumn{4}{l}{\textit{2. Achieved Media Throughput}} \\
        Uplink throughput / publisher (Mbps) & 0.81 & 0.81 & 0.81 \\
        Total uplink throughput (Mbps) & 0.81 & 0.81 & 0.81 \\
        Downlink viewer (mean Mbps) & 0.81 & 0.81 & 0.81 \\
        Total downlink throughput (Mbps) & 50.85 & 50.83 & 50.95 \\
        Rx/Tx fan-out delivery ratio & 100\% & 99.3\% & 99.3\% \\
        \hline
        \multicolumn{4}{l}{\textit{3. End-to-End Latency \& Reliability}} \\
        RTP transport $P_{50}$ (ms) & 52.65 & 267.47 & 230.02 \\
        RTP transport $P_{95}$ (ms) & 128.86 & 699.92 & 617.10 \\
        RTP transport $P_{99}$ (ms) & 155.66 & 810.45 & 712.30 \\
        RTP packet loss & 0.00\% & 0.22\% & 0.00\% \\
        Frame loss & 0.00\% & 0.50\% & 0.06\% \\
        \hline
    \end{tabular}%
    }
\end{table}

\subsection{MCData Service}
\begin{figure}
    \centering
    \includegraphics[width=\columnwidth]{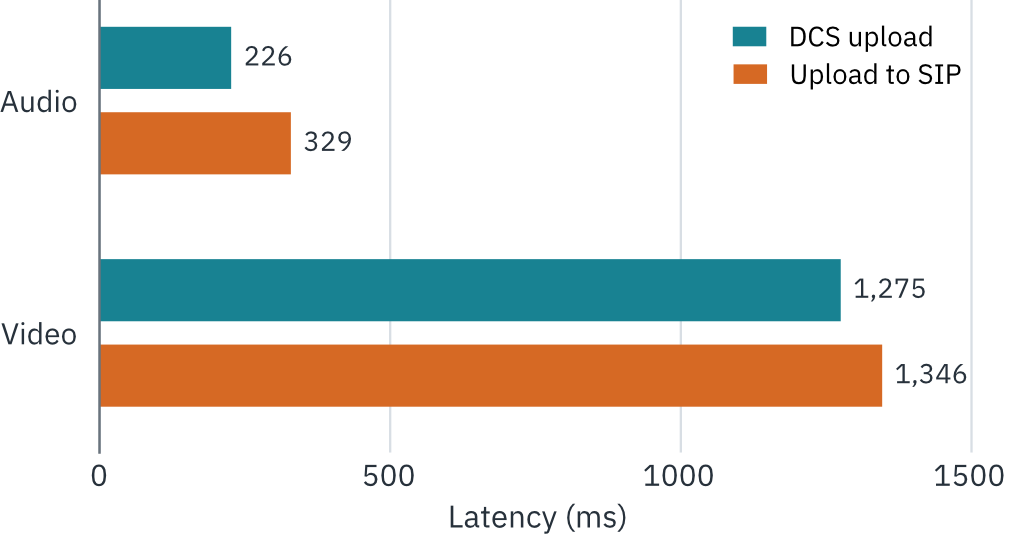}
    \caption{$P_{95}$ \ac{dcs} upload time and upload to \ac{sip} notification latency for audio and video attachments. UE downloads are excluded.}
    \label{fig:mcdata-dcs-latency}
\end{figure}

\ac{mcdata} group messaging was evaluated with one sender and 63 recipients. In each payload run, the sender transmitted 1,000 messages at a rate of 1~Hz. The telemetry and short data messages had 256~B \ac{sip} MESSAGE bodies. The attachment runs used a 6 second, 57,577~B audio clip and a 15 second, 3,029,970~B video clip.
For each attachment, the sender uploaded the file to the \acl{dcs} and sent its 32~B content identifier in a \ac{sip} MESSAGE. Figure~\ref{fig:mcdata-dcs-latency} separates the \ac{dcs} upload time from the time between upload initiation and receipt of the \ac{sip} notification. All 1000 uploads completed in each attachment run. At the 95th percentile, audio uploads took 226.48~ms and video uploads took 1,275.03~ms. The corresponding upload-to-notification times were 329.18~ms and 1,346.33~ms. These values characterize the tested reference workload, and do not include recipient download or playback.



\begin{figure}
    \centering
    \includegraphics[width=\columnwidth]{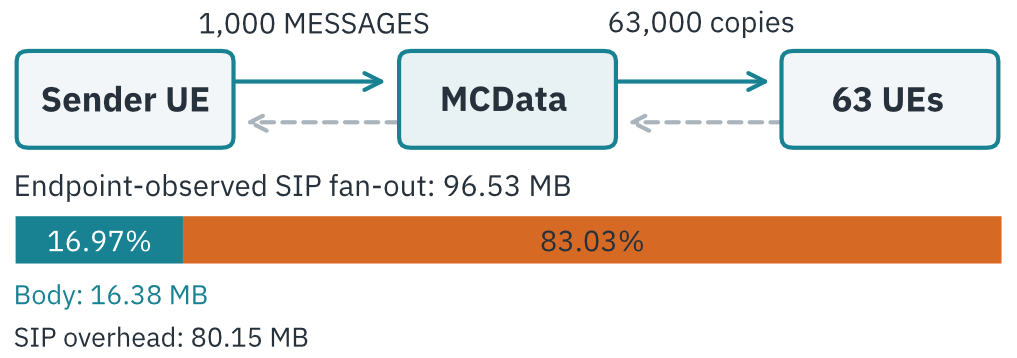}
    \caption{Telemetry \ac{sip} fan-out and endpoint-observed byte accounting. Dashed arrows denote SIP responses. TCP/IP and internal server traffic are excluded.}
    \label{fig:mcdata-signaling}
\end{figure}

Figure~\ref{fig:mcdata-signaling} illustrates the telemetry message fan-out. The sender issued 1,000 \ac{sip} MESSAGE requests, which produced 63,000 recipient messages. Solid arrows show MESSAGE traffic and dashed arrows show \ac{sip} responses. During this exchange, the endpoints recorded 96,531,634 \ac{sip} bytes: 16,384,000 body bytes (16.97\%) and 80,147,634 non-body \ac{sip} bytes (83.03\%), including headers and responses. This byte accounting applies to the telemetry run and excludes TCP/IP framing, internal server traffic, and \ac{dcs} transfers.

To examine the effect of increasing the sending rate, a separate telemetry sweep test with 250 messages, at each of four rates, was conducted. Table~\ref{tab:mcdata-scalability} reports the resulting recipient delivery rate, end-to-end 99th-percentile latency, and successful delivery ratio. All 63,000 expected recipient deliveries were observed across the sweep. As the sending rate rose from 0.5 to 2.0 messages/s, the delivery rate rose from 31.50 to 126.00 deliveries/s, while $P_{99}$ latency increased from 67.14 to 79.07~ms. The 100\% delivery ratio describes these measured runs.

\begin{table}
    \centering
    \caption{Telemetry rate-sweep campaign with 64 \ac{ue}s and 1000 messages
    in total (250 per rate). Send and delivery rates are in messages/s;
    $P_{99}$ is end-to-end delivery latency in ms.}
    \label{tab:mcdata-scalability}
    \small
    \begin{tabular*}{\columnwidth}{@{\extracolsep{\fill}}lcccc@{}}
        \hline
        \textbf{Target} & \textbf{Send rate} & \textbf{Delivery rate} &
        \textbf{$P_{99}$} & \textbf{Success \%} \\
        \hline
        0.5 & 0.50 & 31.50  & 67.14 & 100.00\% \\
        1.0 & 1.00 & 63.00  & 67.54 & 100.00\% \\
        1.5 & 1.50 & 94.50  & 75.05 & 100.00\% \\
        2.0 & 2.00 & 126.00 & 79.07 & 100.00\% \\
        \hline
    \end{tabular*}
\end{table}

\subsection{Discussion of Experimental Findings}
The findings illustrate the utility of distinguishing mission-critical media delivery from service control across a tactical edge 3D network. While the \ac{upf} and RTPengine were centralized across the satellite backhaul (S2), the \ac{mcptt} audio transport delay $P_{95}$ reached 117.53~ms, and packet loss reached 9.78\%. However, by placing these functions at the tactical edge (S3), the corresponding values were reduced to 57.50~ms and 0.36\%, respectively. In addition, audio delivery in S3 was close to that obtained over the terrestrial connection (S1), where the corresponding values were 60.64~ms and 0.73\%. These results demonstrate the benefit of tactical edge function placement within a 3D public-safety network: latency-sensitive media can remain within the local tactical network, while satellite connectivity is used to reach centralized service and control functions.

The platform also supported \ac{mcptt} service operation with up to 64 simulated \acp{ue} across all evaluated configurations. The protocol-level \ac{mcptt} access time $P_{95}$ in S3 was 184.52~ms, remaining below the 300~ms reference for established calls. Furthermore, the higher access time in S3 compared with S2 illustrates that media placement and service access are affected by different components of the end-to-end path: the media relay was local to the tactical edge in S3, whereas \ac{ims} and service-control functions remained centralized and their signaling continued to traverse the \ac{ntn} backhaul. The \ac{mcdata} experiments additionally demonstrated reliable \ac{sip} message distribution at the examined rates, with all 63,000 expected recipient deliveries observed in each reported payload run. These results reinforce the importance of evaluating the complete service chain across the distributed terrestrial, aerial, and non-terrestrial network segments with tactical edge processing rather than considering network transport in isolation.

The controlled contention experiments provide evidence of the proposed \ac{qpp} mechanism under constrained radio resources. In these experiments, the Quality component remained fixed through the service-specific \ac{5qi}~65, while the Priority and Preemption components were varied through the corresponding \ac{arp} profiles. $UE_3$ preempted $UE_4$'s normal-priority dedicated flow, whereas $UE_4$ retained its flow against the same contender under permanent emergency priority. In the dynamic run, two distinct senders with the same $UE_3$ profile requested resources during and after emergency protection. The first was refused a dedicated flow while $UE_4$ was protected; the second was admitted and preempted $UE_4$ after normal priority was restored. Both policy updates modified the existing flow while preserving \ac{5qi}~65. A separate UE-to-RAN experiment measured completion of the \ac{mcx}-initiated QPP control path through \ac{ims}, the \ac{as}, and the 5G Core. The two evaluations, therefore, provide complementary evidence of QPP operation: admission and preemption outcomes under contention, and signaling latency for service-triggered policy updates across terrestrial and satellite-backed deployments.

\section{Conclusions and Future Work}
\label{sec:conclusions}

Resilient public-safety communications require network architectures that can adapt both to changing operational priorities and to the availability of communication infrastructure. This work addresses these requirements through an integrated \ac{mcx} platform combining dynamic \acf{qpp} enforcement with distributed edge processing across terrestrial, aerial, and non-terrestrial 5G connectivity. The platform supports \ac{mcptt}, \ac{mcvideo}, and \ac{mcdata}, while translating operational priority changes into dynamic \ac{arp} updates on active \ac{qos} flows without modifying the service-specific \ac{5qi} or disrupting established sessions.

Through experimental evaluation with up to 64 simulated \acp{ue}, the proposed solution demonstrated the benefits of tactical-edge media processing under satellite backhaul. Compared with centralized media processing, edge placement reduced \ac{mcptt} audio transport delay $P_{95}$ from 117.53 ms to 57.50 ms and packet loss from 9.78\% to 0.36\%, approaching the terrestrial baseline. The edge configuration further achieved an \ac{mcptt} access-time $P_{95}$ of 184.52 ms and reliable \ac{mcdata} \ac{sip} delivery at the evaluated message rates. These results demonstrate that local media processing can preserve responsive mission-critical services while reducing their dependence on constrained non-terrestrial backhaul.

Together, the \ac{qpp} and tactical edge results provide an experimental foundation for context-aware \ac{mcx} operation across 3D public-safety networks. The contention tests demonstrate profile-dependent admission and retention with unchanged service-specific \ac{5qi}, while the placement experiments show the benefit of keeping local \ac{mcptt} media at the tactical edge. Future work will extend the priority evaluation to repeated trials and a wider range of competing loads, and examine service recovery after backhaul interruption and \ac{mcvideo} performance during aerial operation.

\section*{Acknowledgment}

This work was supported by the European Regional Development Fund (ERDF), through the Centro Regional Programme (CENTRO 2030) of the Portugal 2030 framework [Project DAWN6G with Nr. 19219 (CENTRO2030-FEDER-01672300)], and through the Innovation and Digital Transition Programme (COMPETE 2030) of the Portugal 2030 framework [Project ALBATROZ with Nr. 17775 (COMPETE2030-FEDER-01335900)].

\bibliographystyle{IEEEtran}
\bibliography{bib/references}

\end{document}